\documentclass[aps,groupedaddress,superscriptaddress,showpacs,prb,floatfix]{revtex4}

\usepackage{graphicx}
\usepackage{dcolumn}
\usepackage{bm}
\usepackage{epsfig}
\usepackage{dsfont,psfrag}
\usepackage{color}
\newcommand{\be}{\begin{eqnarray}}
\newcommand{\ee}{\end{eqnarray}}
\def\refeq#1{(\ref{#1})}
\renewcommand{\vec}[1]{\boldsymbol #1}
\def\d{\mbox d}
\def\wt{\widetilde}
\def\nn{\nonumber}

\def\o{\omega}

\def\L{\Lambda}

\def\Or{\mathcal O}
\def\g{\gamma}

\def\ve{\varepsilon}

\def\l{\left}
\def\r{\right}

\def\rmi{{\rm i}}

\def\up{\uparrow}
\def\Up{\Uparrow}
\def\down{\downarrow}
\def\Down{\Downarrow}
\def\M{\mathcal{M}}
\def\ov{\overline}

\begin{document}
\bibliographystyle{apsrev}
  \title{Spectrum and screening cloud in the central spin model}
\author{Michael Bortz}
\email{bortz@physik.uni-kl.de}
\affiliation{Fachbereich Physik und Research Center OPTIMAS, Technische Universit\"at Kaiserslautern, Erwin-Schr\"odinger-Str., 67663 Kaiserslautern, Germany}
\author{Sebastian Eggert}
\affiliation{Fachbereich Physik und Research Center OPTIMAS, Technische Universit\"at Kaiserslautern, Erwin-Schr\"odinger-Str., 67663 Kaiserslautern, Germany}
\author{Joachim Stolze}
\affiliation{Institut f\"ur Physik, Technische Universit\"at Dortmund, 44221 Dortmund, Germany}
\pagestyle{plain}

\date{\today}
\begin{abstract}
We consider an electronic spin in a quantum dot, coupled to the surrounding
nuclear spins via inhomogeneous antiferromagnetic hyperfine interactions and subject to a uniform field,
{which is described by Gaudin's central spin model}.
We study spectral properties, 
the two-point
correlation functions, and the magnetization profile in the ground state and in low-lying excited states,
which characterizes the structure of the cloud of nuclear spins screening the electron spin. 
A close connection to the pair occupation probability in the BCS-model is established. Using
the exact Bethe Ansatz solution of that
  model and arguments of integrability, 
we can distinguish between contributions from purely classical physics and 
from quantum fluctuations. 
\end{abstract}
\pacs{73.21.La,02.30.Ik}
\maketitle

\section{Introduction}
Over the last decade, experimental realizations of strongly correlated quantum
systems have led to the possibility of studying non-equilibrium quantum processes on
a microscopic level. From a theoretical point of view, the description of such
processes is most challenging because it requires a thorough study of
the spectrum and correlation functions. 

In this work, we consider a model which describes the hyperfine interaction
{of} an electron spin (the central spin) 
in a quantum dot with {a bath of} nuclear spins
in the dot.
The resulting Heisenberg exchange interaction is dominant for
short time scales up to 1ms\cite{joh05} before {other} mechanisms like spin-orbit coupling or dipole-dipole-interactions between the bath spins set in.\cite{kha00,kha01,sch03} This is an ideal system to {generally}
understand the decoherence of a qubit which is realized by the electron spin,\cite{los98} and in this context the loss of quantum information. Many important
contributions on this central issue have been made by a number of authors using
different
methods,\cite{kha02,kha03,sch02,sch03,dob03,coi04,erl04,sem03,den06,has06,coi06,zha06,che07,kur09,mer09} as also outlined in the reviews [\onlinecite{han07,coi09}].
All those works rely on often very sophisticated approximation schemes
to study the time evolution of the central spin directly. In this work, our goal is to use the 
{\it exact} solution of the model to study its spectrum and static correlation
functions in the ground state and excited states
also in comparison with a simple classical approximation.
In the future, this knowledge can be used to obtain exact information about non-equilibrium
dynamics like the decoherence process. 

The central spin model (or Gaudin model\cite{gau76,gaubook}) 
we consider here describes the
isotropic Heisenberg {coupling} of the central electron spin $\vec S_0$
with inhomogeneous exchange couplings $A_j$ to a bath of $N_b$ nuclear spins $\vec S_{j=1,\ldots,N_b}$. The nuclei are assumed 
to be spin-1/2 particles and their coupling $g_{\rm n}$ to the external magnetic 
field $h$ is assumed to be much weaker than that of the electron, $g_{\rm e}$,
\be
H=\sum_{j=1}^{N_b}A_j \vec S_0 \cdot \vec S_j - h g_{\rm e} S_0^z-h g_{\rm n} \sum_{j=1}^{N_b} S_j^z\label{gaudef}\,.
\ee 
The couplings $A_j$ are proportional to the square of the electronic wave
function at the positions of the nuclei. For a realistic distribution of the $A_j$, we can think of the index $j$ as
measuring the distance from the center of the dot. 
{The methods we use in this work, especially the classical approach and the integrability, do not depend on the choice 
of couplings $A_j$, but for definiteness we assume a 
harmonic trapping potential for the electron.
This results in a Gaussian decay of the couplings\cite{coi04} 
\be
A_j=\alpha \exp\l[-( jB/N^{1/D}_b)^2\r]\,, \label{coup}
\ee
where the normalization $\alpha=x_1 N_b/\sum_{j=1}^{N_b}\exp\l[-( jB/N^{1/D}_b)^2\r]$ is chosen
such that the 
mean value (or first moment) $x_1$ of the $A_j$ is fixed and the dimension is taken $D=1$.}
Here, the parameter $B$ controls the degree of inhomogeneity. We will choose
$B=2$, $x_1=2$ as generic values for inhomogeneous couplings and
$B=2/5$, $x_1=2$ as an example for nearly homogeneous couplings in {numerical diagonalizations in later sections}. 

{We calculate} the spectrum, the magnetization profile 
$\langle S_j^z\rangle$ {of the nuclear bath spins},
and the two-point functions $\langle \vec S_0\cdot \vec
S_j\rangle$ of the model \refeq{gaudef}. {It is possible to 
distinguish two types of contributions in these quantities: }
On the one hand, terms appear that can be
obtained from a purely classical approach. Additionally, we identify terms
stemming from quantum fluctuations. Most importantly, classical and quantum
terms can be of the same order in the two-point function.    

This paper is organized as follows. In Sec.~\ref{sec:gen}, we show how to obtain
one- and two-point functions from the exact Bethe Ansatz solution 
for the eigenvalues and eigenstates of the Hamiltonian \refeq{gaudef}. The central spin and BCS pairing models are linked by their integrability which provides a way to calculate the 
magnetization profile $\langle S_j^z\rangle$. 
Two-point functions are given by derivatives of the energy with respect to the 
Heisenberg coupling constants. 

In the third section, we evaluate one- and two-point correlation functions
based on a classical picture which, for finite magnetic field, assumes
spontaneous symmetry breaking in the model \refeq{gaudef}, similar to the
superconducting phase transition in the {closely related} BCS model. The local magnetization
obtained by completely diagonalizing the quantum mechanical model with 16 spins agrees very
well with the classical results. However, for the two-point function, the
agreement is less good, which indicates that quantum fluctuations are of the
same order as the classical terms. 
 
We are thus led to study the exact solution in Sec.~\ref{sec:corr}, especially in
order to obtain quantum mechanical contributions to correlation
functions. This is done for zero and finite magnetic fields. The connection
with the classical approach 
is also established. The
paper ends with an outlook.


\section{Exact solution, link to the BCS model and correlation functions}
\label{sec:gen}
\subsection{Exact solution}
With a special focus on the magnetic field terms, we rewrite Eq.~\refeq{gaudef} as
\be
H=\sum_{j=1}^{N_b}A_j \vec S_0\cdot \vec S_j - h_0 S_0^z-h_{\rm t}S^z_{\rm tot}\label{ham}\,
\ee 
with $h_0=h(g_{\rm e}-g_{\rm n})$, $h_{\rm t}=h g_{\rm n}$ and the total polarization $S^z_{\rm tot}=\sum_{j=0}^{N_b} S_j^z=\frac{N}{2}-M$, where $N=N_b+1$ is the total number of spins and $M$ is the number of flipped spins compared to the ferromagnetic all-up state. 
Note that $S^z_{\rm tot}$ commutes with the Hamiltonian,
\be
\l[H,S^z_{\rm tot}\r]=0\label{commut},
\ee
and thus $S^z_{\rm tot}=N/2-M$ is a constant of motion. This means that the last term in (\ref{ham}) provides an additive constant which we will drop in the following unless otherwise stated.

The model \refeq{ham} has been solved by Gaudin \cite{gau76,gaubook} using a coordinate-type Bethe ansatz; an algebraic solution has been given by Sklyanin,\cite{skl89} and is also described in Ref.~[\onlinecite{del02}]. The exact solution has been used in Ref.~[\onlinecite{bor_inhom07}] to calculate non-equilibrium dynamics in a fully polarized bath. Using the notation from Ref.~[\onlinecite{bor_inhom07}], the eigenvalues $\L$ in a sector of given $M$ read
\be
\L&=&-\frac12 \sum_{k=0}^{M_b}\omega_{k}+\frac{N_b x_1}{4}-\frac{h_0}{2}\,\;\;,\label{ev}
\ee
{where $x_1$ is the mean value of the $A_j$ and $M_b:=M-1$. The set of the $\omega_k$, $k=0,\ldots,M_b$, is determined by the Bethe Ansatz equations (BAE)}
\be
1+\sum_{j=1}^{N_b}\frac{A_j}{A_j-\omega_{k}}-2 \sum_{k'\neq k}^{M_b}\frac{\o_{k'}}{\o_{k'}-\o_{k}}+\frac{2 h_0}{\o_{k}}=0\label{bae}\,.
\ee
Gaudin \cite{gau95} showed that there are $C^N_M=N!/(M!(N-M)!)$ sets of solutions $\l\{\o_{0},\ldots,\o_{M_b}\r\}$ 
to these equations in each sector of given $M$, one for each eigenvalue $\Lambda$. 
The corresponding energy eigenstates with a fixed number $M$ of flipped spins are given by
\be
 |M\rangle&=&\frac{1}{n_{M}}\prod_{k=0}^{M_b}\l[-S_0^-+\sum_{j=1}^{N_b}\frac{A_j}{\o_{k}-A_{j}}S_j^-\r]|0\rangle\label{baes},
\ee
where $|0\rangle$ is the fully polarized state $|\Up;\up,\ldots,\up\rangle$, and the arrows $\Up,\Down$ for the central spin and $\up,\down$ for the bath spins are used. The normalization factor $n_{M}$ was conjectured by Gaudin \cite{gaubook,gau95} and proved by Sklyanin \cite{skl99} for $h_0=0$: 
\be
n^2_{M}&=&(-1)^{M} \det \M\nn\\
\M_{kk}&=&-1-\sum_{j=1}^{N_b}\frac{A_j^2}{(\o_{k}-A_j)^2}+\sum_{k'\neq k}\frac{2 \,\o_{k'}^2}{(\o_{k}-\o_{k'})^2}\nn\\
\M_{kk'}&=&-\frac{2 \,\o_{k'}^2}{(\o_{k}-\o_{k'})^2},\; k\neq k'\nn.
\ee
In Ref.~[\onlinecite{bor_inhom07}] evidence was given that this holds for finite $h_0$ as well.

Let us now come back to the eigenvalues in Eq.~\refeq{ev}. Due to the Hellmann-Feynman theorem,\cite{gue32,pau33,hel37,fey39} two-point correlators between the central spin and a bath spin in an eigenstate are obtained as the derivatives of the energy eigenvalues
\be
\langle \vec S_0\cdot \vec S_j\rangle=\partial_{A_j} \L\label{2point},
\ee
and the expectation value of the central spin polarization {is given by}
\be
\langle S_0^z\rangle =-\partial_{h_0} \L.
\ee
{By solving the BAE \refeq{bae} as a function of the couplings $A_j$ and the 
field $h_0$ it is therefore possible to obtain the expectation values directly.}

{In order to also calculate the magnetization profile 
$\langle S_j^z\rangle$, $j=1,\ldots,N_b$ we have to use some
additional features of the integrable structure of the model \refeq{ham}, 
as will be described in the remainder of this section. }
Let us rewrite Eq.~\refeq{ham} in the original notation used by Gaudin,\cite{gaubook} 
\be
H_\ell=-\sum_{j=0,j\neq \ell}^{N_b} \frac{\vec S_\ell\cdot \vec S_j}{\ve_\ell-\ve_j}-h_0 S_\ell^z,\label{hell}
\ee
such that we recover Eq.~\refeq{ham} with $h_{\rm t}=0$ for 
\be 
A_j=1/\ve_j\label{coup_eq}
\ee
and $\ell=0$, $\ve_0=0$ in Eq.~\refeq{hell}. As pointed out by Gaudin, \cite{gaubook} 
\be
\l[H_\ell,H_{\ell'}\r]=0
\ee
which means that an integrable Hamiltonian can be constructed as a linear combination of $N$ mutually commuting conserved quantities
\be
\widetilde H:=\sum_{\ell=0}^{N_b} \ve_\ell H_\ell = -h_0 \sum_{\ell=0}^{N_b} \ve_\ell S_\ell^z - \frac12 \l(\vec S_{\rm tot}\r)^2+ \frac12 \sum_{\ell=0}^{N_b} \vec S_\ell^2\label{tham},
\ee
with $\vec S_{\rm tot}=\sum_{\ell=0}^{N_b} \vec S_\ell$. 

{The model \refeq{tham} has the same eigenstates as the original model \refeq{ham}, even though these are not necessarily in the same energetic order. For the local expectation values,} one can apply the Hellmann-Feynman theorem\cite{gue32,pau33,hel37,fey39} to the eigenvalues $\widetilde \Lambda$ of $\widetilde H$: 
\be
\langle S_j^z\rangle=-\frac{1}{h_0}\partial_{\ve_j} \widetilde \Lambda=-\frac{1}{h_0}\partial_{A_j^{-1}} \widetilde \Lambda\label{sjz1}.
\ee
In order to calculate $\widetilde \Lambda$, we use Gaudin's result\cite{gau76} for the eigenvalues $\Lambda^{(\ell)}$ of $H_\ell$ in Eq.~\refeq{hell}
\be
\Lambda^{(\ell)}=\frac{1}{2} \sum_{k=0}^{M_b}\frac{1}{\ve_\ell-E_{k}}-\frac14 \sum_{j=0,j\neq\ell}^{N_b} \frac{1}{\ve_\ell-\ve_j} - \frac{h_0}{2}
\ee
with 
\be
E_{k}=1/\omega_{k},\label{ekok}
\ee
such that
\be
\widetilde \Lambda&=&\sum_{\ell=0}^{N_b} \ve_\ell \Lambda^{(\ell)}\nn\\
&=& \frac12 \sum_{\ell,k} \frac{\ve_\ell}{\ve_\ell-E_{k}} - \frac{N_b(N_b+1)}{8}-\frac{h_0}{2} \sum_{\ell=0}^{N_b} \ve_\ell\label{wtlambdanu}\,.
\ee
By rewriting the BAE \refeq{bae} in terms of the $\ve_j$, $E_{k}$ and defining $g:= 1/h_0$, we arrive at
\be
\sum_{j=0}^{N_b}\frac{1}{E_{k}-\ve_j}-2\sum_{k'\neq k}\frac{1}{E_{k}-E_{k'}}+\frac{2}{g}=0. \label{auxeq}
\ee
Observing that $\varepsilon_j/(E_{k}-\varepsilon_j) = E_{k}/(E_{k}-\varepsilon_j)-1$ and $ E_{k}/(E_{k}- E_{k^{\prime}})  +  E_{k^{\prime}}/(E_{k^{\prime}}- E_{k}) =1$ we can eliminate the first term in Eq.~\refeq{wtlambdanu}
{by multiplying \refeq{auxeq} with $E_{k}$ and then performing the sum over $k$. 
Hence Eq.~(\ref{wtlambdanu}) becomes}  
\be
\widetilde \Lambda&=&h_0 \sum_{k=0}^{M_b} E_k - \frac{h_0}{2} \sum_{\ell=1}^{N_b} \ve_\ell -\frac{M_b(M_b+1)}{4}-\frac{N_b(N_b+1)}{8}+\frac{(N_b+1)(M_b+1)}{2}\label{lambdanu},
\ee
which yields, together with Eq.~\refeq{sjz1},
\be
\langle S_j^z\rangle&=&\frac12-\partial_{\ve_j}\sum_{k=0}^{M_b} E_k=\frac12-\partial_{1/A_j}\sum_{k=0}^{M_b} \frac{1}{\omega_k}\label{sjznu}.
\ee
{In summary it is therefore possible to express the 
two-point function 
\refeq{2point} and the local magnetization \refeq{sjznu}
in terms of the BA numbers of the exact solution, which is the main finding of this section.  These quantities 
will be analyzed in detail in sections \ref{tl} and \ref{sec:corr}.}

\subsection{Link to the BCS pairing model}
{It is possible to relate spin with fermionic operators, 
using Anderson spin-1/2 pseudospin operators\cite{and58,del02}
\be
S^z_j&=&\frac12\l(1-c^\dagger_{j\up}c_{j\up} -c^\dagger_{j\down}c_{j\down}\r)\label{ps1},\\
S^-_j&=& c^\dagger_{j\up}c^\dagger_{j\down}\;,\qquad S^+_j= c_{j\down} c_{j\up}\label{ps2}\;,
\ee
which preserve the $SU(2)$ commutators $\l[S_i^+,S_j^-\r]=2\delta_{ij} S_j^z$, $\l[S_i^z,S_j^{\pm}\r]=\pm \delta_{ij}S_j^\pm$.\cite{del02} }

{A BCS-like Hamiltonian can be defined by rescaling the integrable model  $\widetilde H$ from Eq.~\refeq{tham} 
\be
H_{\rm BCS}:=\frac{1}{h_0} \widetilde H + \frac{1}{2} \sum_{\ell=0}^{N_b} \ve_\ell+\frac{S^z_{\rm tot}\l(S^z_{\rm tot}+1\r)}{2h_0} -\frac{3(N_b+1)}{8 h_0}\label{prop},
\ee
where $S^z_{\rm tot}=N_b/2-M_b-1/2$ is the conserved quantum number from Eq.~(\ref{ham}). 
In terms of spin operators $H_{\rm BCS}$ therefore reads}
\be
H_{\rm BCS}=-\sum_{j=0}^{N_b} A_j^{-1} \l(S_j^z-\frac12\r) - \frac{1}{2h_0} (\vec S_{\rm tot})^2 + \frac{1}{2h_0} S^z_{\rm tot} \l(S^z_{\rm tot}+1\r)\;.
\ee
Replacing the spin-operators through Eqs.~\refeq{ps1} and \refeq{ps2}, one arrives at the fermionic representation 
\be
H_{\rm BCS}=\frac{1}{2}\sum_{{\ell=0\atop \sigma=\up,\down}}^{N_b} \ve_\ell c^\dagger_{\ell\sigma} c_{\ell\sigma}- \frac{g}{2} \sum_{\ell,j=0}^{N_b}c^\dagger_{\ell\down}c^\dagger_{\ell\up}c_{j\up} c_{j\down}\label{hambcs},
\ee
with the doubly degenerate single-particle levels $\ve_\ell$ and the pairing amplitude $g=1/h_0$.\cite{cam97} 

The Hamiltonian \refeq{hambcs} describes $M$ pairs of fermionic particles interacting via an attractive pairing potential, thus affecting the $N$ doubly degenerate energy levels $\ve_j$. In a series of papers, Richardson \cite{ric62,ric63,ric64a,ric64b,ric65,ric77} used it to describe pairing in nuclei. In the more recent past, the exact solution of this model has been rediscovered to study ultrasmall metallic grains in their superconducting phase.\cite{del01} In the thermodynamic limit, the solution of the model \refeq{prop} yields the mean-field BCS solution;\cite{bcs57} we will come back to this point in Sec.~\ref{tl}. 

From Eqs.~\refeq{sjz1} and \refeq{prop} it follows that the occupation probability 
$\langle n_j\rangle:=\langle c^\dagger_{j\up}c_{j\up} +c^\dagger_{j\down}c_{j\down} \rangle/2$ 
of the single particle level $\ve_j$ reads \cite{ric77,rom02} 
\be
\langle n_j\rangle&=&\partial_{\ve_j}\langle H_{\rm BCS}\rangle \nn\\
&=&\frac12-\langle S_j^z\rangle  \label{rel},
\ee    
which is consistent with Eq.~\refeq{ps1}. 
Thus the single particle occupation numbers in the pairing model are directly related to the local polarization of nuclear spins in the central spin model. 

We will compute two-point correlation functions and the magnetization profile for different parameter regimes in the following sections. For illustrative purposes, let us first check the extreme limits $h_0\to 0,\infty$ in Eq.~\refeq{rel} for the ground state in the sector $S^z_{\rm tot}=0$ (this implies that we take $N$ to be even here). In the BCS-model, this corresponds to the case of half filling, where the number of electrons $2M$ equals the number of free particle levels $N$. For $h_0\to 0$, the model \refeq{ham} is $SU(2)$-invariant, so $\langle S_j^z\rangle|_{h_0=0}=0$. Since $g=1/h_0$, the pairing potential is infinitely strong in this limit, such that all levels are occupied and only ideal Cooper pairs exist, where each level is occupied by half a pair. 

\begin{figure}
\psfrag{j}{$j$}
\psfrag{m}{$\langle S_j^z\rangle$}
\psfrag{a}{$A_j$}
\psfrag{e}{$\varepsilon_j$}
\psfrag{n}{$\langle n_j\rangle$}
\includegraphics[width=0.7\textwidth]{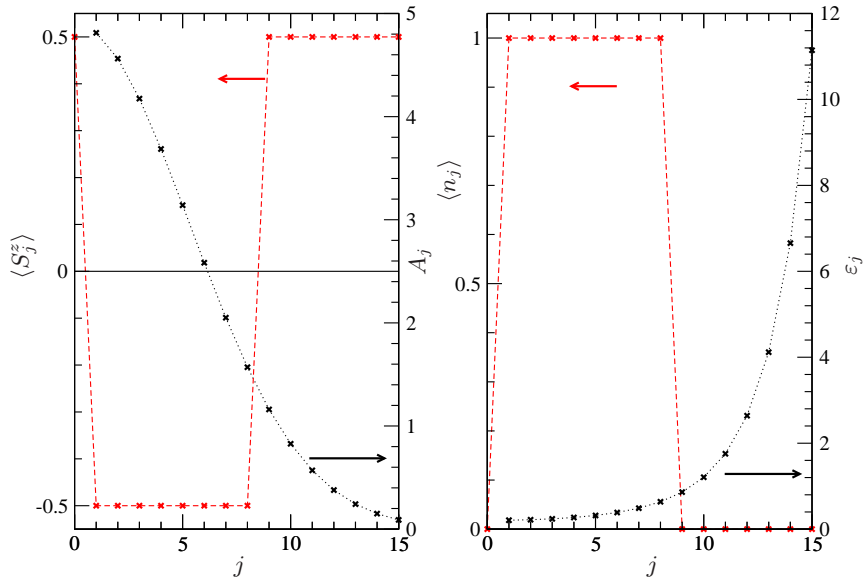}
\caption{{(color online) Left panel: Magnetization profile in a quantum dot with $N_b=15$, infinite central magnetic field, 
$h_0\to \infty$, and total field $A_{8}>2h_t>A_{9}$. The black crosses denote the coupling constants chosen according to 
Eq.~\refeq{coup} with $x_1=2,\,B=2$. Right panel: Corresponding electronic occupation probability according 
to Eq.~\refeq{rel} for free Fermions, with (black crosses) the single-particle levels $\ve_j$, $j=1,\ldots,N_b$, 
according to Eqs.~\refeq{coup} and \refeq{coup_eq}. The system is in a two-particle excited state, 
where the pair occupying the lowest energy level $\ve_0=0$ is shifted to the energetically lowest 
state above the Fermi level.}}
\label{fig:mag_prof_infinite_field}
\end{figure}

In the opposite limit, $h_0\to \infty$, the central spin 
is frozen along the $z$-direction. {The directions of the bath spins
are simply given by the competition of the antiferromagnetic exchange 
in Eq.~\refeq{ham} with  
the magnetic field $h_{\rm t}$.  
Therefore, all outer bath spins with coupling $A_j<2h_{\rm t}$ are aligned
with the field and the central spin, while the inner ones point in the opposite 
direction. The resulting magnetization profile is sketched schematically 
in the left panel of 
Fig.~\ref{fig:mag_prof_infinite_field}, where 
we chose $A_{(N_b+1)/2}>2h_t>A_{(N_b+1)/2+1}$, such that $S^z_{\rm tot}=0$ for illustrative purposes. }
 
For the BCS model \refeq{hambcs} this means that the highest states where $\ve_j$ is largest (i.e. $A_j=1/\ve_j$ is smallest) are unoccupied. This is the filled Fermi sea for the non-interacting Fermi gas. The level $\ve_0=0$ is special in the sense that it is unoccupied in the ground state of the central spin model, which is an excited state in terms of the BCS Hamiltonian. From this we conclude that the ground state of the central spin model for $S^z_{\rm tot}=0$ corresponds to an excited state of the BCS model where the energetically lowest pair is shifted to the top of the filled Fermi sea. This is illustrated in the right panel of Fig.~\ref{fig:mag_prof_infinite_field} and will be further discussed in Sec.~\ref{bcs}. 


\section{The screening cloud from a classical point of view}
\label{tl}
In this section, we develop a classical picture for the energy and the magnetization profile of the model \refeq{ham} for finite magnetic fields, which turns out to be closely related to the mean-field BCS solution \cite{bcs57} of the pairing Hamiltonian \refeq{hambcs}. 

It is reasonable to expect that for large coordination number $N_b\gg 1$, a classical approach to the Hamiltonian \refeq{ham} yields valuable insights into the physics of the model.\cite{yuz05} The classical approach consists of replacing quantum-mechanical spin operators $\vec S_j$ by classical vectors $\langle \vec m_j\rangle$. Especially, for states with the same quantum number $S^z_{\rm tot}$, an expectation value $\langle S_j^x\rangle\neq 0$ implies that in this limit, the Hamiltonian symmetry \refeq{commut} is spontaneously broken. This mechanism is analogous to the superconducting phase transition in which particle number conservation is broken, $\langle c^\dagger_{j\up}c^\dagger_{j'\down}\rangle\neq 0$. 

\subsection{Magnetization pattern in the central spin model}
Let us begin by parameterizing each spin in polar coordinates, $\vec m_j=\frac12(\cos \varphi_j\,\sin \vartheta_j,\sin \varphi_j\,\sin \vartheta_j,\cos \vartheta_j)$ such that $|\vec m_j|^2=\frac{1}{4}$ for $j=0,\ldots,N_b$. Our aim is to derive the ground state configuration described by the angles $\varphi_j, \,\vartheta_j$ for a given total magnetization $S^z_{\rm tot}$ and fixed central field $h_0$. 

\begin{figure}
\psfrag{a}{$\vartheta_0$}
\psfrag{b}{$\vartheta_1$}
\psfrag{f}{$\vartheta_{N_b}$}
\includegraphics[scale=0.5]{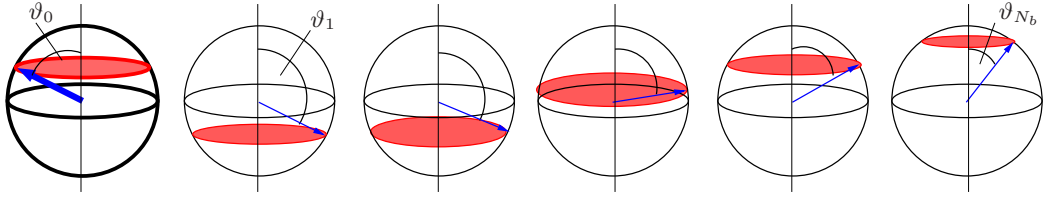}
\caption{(color online) Schematical orientation of the classical spins according to Eqs.~\refeq{ans1} and \refeq{ans2}. The fat leftmost
  spin is the center of the dot. The central field leads to a canting of the central spin, which, due to the antiferromagnetic exchange, leads to an opposite canting of the neighboring spins. Since an overall magnetic field is included which fixes the total magnetization, a non-trivial magnetization profile results.}
\label{fig:clas}
\end{figure}

{The classical energy as a function of the azimutal angles $\varphi_j$ is always minimized by choosing 
$\varphi_0-\varphi_j=\pi$, corresponding to antiferromagnetic alignment in the $xy$-plane. 
The resulting classical model for the polar angles analogous to Eq.~\refeq{ham} is 
then given by
\be
H_{\rm cl}= \frac14 \sum_{j=1}^{N_b}A_j \cos\l(\vartheta_0+\vartheta_j\r)-\frac{h_0}{2} \cos \vartheta_0-\frac{h_{\rm t}}{2}\sum_{j=0}^{N_b}\cos \vartheta_j\,,\label{hclangles}
\ee
and the total magnetization can be determined from
\be
2S^z_{\rm tot}=\sum_{j=0}^{N_b} \cos \vartheta_j\,.\label{szangles}
\ee
The first antiferromagnetic term in Eq.~(\ref{hclangles}) is minimized by large polar angles
$\vartheta_0+\vartheta_j=\pi$, i.e.~spins lying in the $xy$-plane, while the field tends to keep
the polar angles small, analogous to the situation in a two-dimensional Heisenberg antiferromagnet 
with a central impurity.\cite{egg07}
For finite fields the central spin typically acquires a relatively small but finite polar angle,
while the bath spins cant into the opposite direction out of the plane with polar angles that 
are closer to $\pi/2$.
Depending on the overall magnetic field this results in a characteristic magnetization profile:
Those bath spins which are coupled strongly are aligned antiferrogmagnetically to the 
central spin (i.e. against the field), while the more loosely bound bath spins at the edge of the dot are aligned ferromagnetically.  
Depending on the parameters the total magnetization $S^z_{\rm tot}$ is often quite small or even negative.
A typical resulting magnetization profile is sketched in Fig.~\ref{fig:clas}. }

{The minimal values of the angles are most easily found by requiring that the total value of 
the torque $|\vec \tau_0|$ experienced by the central spin 
from the central field and the bath spins has to vanish }
\be
|\vec \tau_0|&=&\partial_{\vartheta_0}H_{\rm cl}=0\label{htheta0}\\
\Rightarrow h_0 \sin \vartheta_0&=&\frac12 \sum_{j=1}^{N_b} A_j \sin(\vartheta_j+\vartheta_0)-h_{\rm t} \sin\vartheta_0\;.\label{tau0}
\ee
Equally well, the torque on each individual bath spin is zero in equilibrium 
\be
|\vec \tau_j|&=&\partial_{\vartheta_j}H_{\rm cl}=0\label{hthetaj}\\
\Rightarrow h_{\rm t} \sin \vartheta_j&=&\frac{A_j}{2} \sin(\vartheta_j+\vartheta_0)
{\qquad (j \neq 0)}\;.\label{tauj}
\ee
Obviously, Eqs.~\refeq{tau0} and \refeq{tauj} are trivially fulfilled when $\vartheta_{j=0,\ldots,N_b}$ are multiple integer values of $\pi$. We exclude these solutions here, because generally, 
they do not correspond to minima of the energy, 
as can be seen from the Hesse matrix of second derivatives of $H_{\rm cl}$ . 

We now insert Eqs.~\refeq{tau0} and \refeq{tauj} into Eq.~\refeq{hclangles} and obtain
\be
H_{\rm cl}= -\frac{1}{4} \sum_{j=1}^{N_b} A_j \frac{\sin\vartheta_j}{\sin\vartheta_0}-\frac{1}{4} \sum_{j=1}^{N_b} A_j \cot \vartheta_j \sin(\vartheta_0+\vartheta_j).\label{hclangle2}
\ee
From Eqs.~\refeq{tau0} and \refeq{tauj} it follows that if the fields $h_{0,\rm t}$ are given, then we can solve for the angles ${\vartheta_j}$, which are given by
\be
\tan\vartheta_j=\frac{\delta A_j}{\nu-A_j}\; ,\label{ansangle}
\ee
where
\be
\delta=\tan\vartheta_0, \ \ \ \nu=2 h_{\rm t}/\cos\vartheta_0. \label{ansangle0}
\ee
The angles in Eq.~(\ref{ansangle}) shows the generic behavior described
 above unless extreme values  of the 
parameters are assumed:  The magnetization
changes from alignment with the field for
the outermost bath spins ($A_j\to 0$) through the $xy$-plane ($A_j\sim \nu$) to near antiferromagnetic 
alignment for the most strongly coupled spins near the center ($A_j > \nu$). 

The components of the magnetization along the field $m^z$ and in the plane $m^\perp$ can be found explicitly
by using 
\be
\tan\vartheta_0& = & m_0^\perp/m_0^z =  \delta \,\label{ans1}\\
\tan\vartheta_j & = & m_j^\perp/m_j^z= \frac{\delta\, A_j}{\nu-A_j}\;\;{\rm for}\;\; j=1,\ldots,N_b\label{ans2},
\ee
from which it follows that
\be
m_0^z&=& \frac{1}{2\sqrt{1+\delta^2}} \,,\qquad m_0^\perp=\frac{\delta}{2\sqrt{1+\delta^2}}\label{m0}\,,\\
m_j^z&=& \frac{\nu-A_j}{2\sqrt{(\nu-A_j)^2+(A_j\,\delta)^2}}\,,\qquad m_j^\perp=\frac{\delta A_j}{2\sqrt{(\nu-A_j)^2+(A_j\,\delta)^2}}\label{mj}\; .
\ee
Similar equations were obtained using methods of classical integrability in Ref.~[\onlinecite{yuz05}]. 
{In order to determine the parameters $\delta$ and $\nu$ we obtain from Eqs.~\refeq{m0} and \refeq{mj}
for the total magnetization along the field 
\be
2S^z_{\rm tot}=\frac{1}{\sqrt{1+\delta^2}}+ \sum_{j=1}^{N_b}\frac{\nu-A_j}{\sqrt{(\nu-A_j)^2+(A_j\,\delta)^2}}\equiv 2N-M\label{szmf}\; .
\ee
Equation \refeq{tau0} for the central field now reads
\be
h_0= \sum_{j=1}^{N_b} \frac{\nu A_j}{2\sqrt{(\nu-A_j)^2+(A_j\,\delta)^2}}-\frac{\nu}{2\sqrt{1+\delta^2}}\label{h0mf}.
\ee
Eqs.~\refeq{szmf} and \refeq{h0mf} fix $\delta$ and $ \nu$ uniquely for a given $S^z_{\rm tot}$ and $h_0$, so that 
all classical vectors are known, which is
the central result of this section}. 

Finally, one obtains the corresponding expression 
for the energy from Eq.~\refeq{hclangle2} {without the trivial $h_{\rm t}$-term}
\be
H_{\rm cl}=  -\frac14\sum_{j=1}^{N_b} \l[\frac{1+\delta^2}{(\nu-A_j)^2+(A_j\,\delta)^2}\r]^{1/2}A_j^2\label{hcl}\,.
\ee
This parametrization of the ground state energy in terms of $\nu,\delta$ and the $A_j$ will be helpful in separating classical from pure quantum contributions in the exact solution later on in Sec.~\ref{bcs}. 

It is interesting to note that an alternative derivation of Eq.~\refeq{h0mf} is obtained by considering the magnetic fields $h_{0,\rm t}$ as canonically conjugate to $m_0^z$, $S^z_{\rm tot}$, so that $h_0=\partial_{m_0^z}\sum_{j=1}^{N_b}\vec m_0\cdot \vec m_j$ and $h_{\rm t}=\partial_{S_{\rm tot}^z}\sum_{j=1}^{N_b}\vec m_0\cdot \vec m_j$. 

The classical spin-spin correlation function between the electron and nuclear spins can be obtained from Eqs.~\refeq{m0} and \refeq{mj}, namely
\be
\vec m_0 \cdot \vec m_j&=&-\frac{(1+\delta^2)A_j-\nu}{4\sqrt{1+\delta^2}\sqrt{(\nu-A_j)^2+(A_j\delta)^2}}\label{sosj_class}.
\ee

\subsection{Connection with the BCS-model}
Very similar relations were derived \cite{bcs57,gau76,ric77} for the thermodynamic limit of the 
BCS-pairing model \refeq{hambcs}
\be
H_{\rm BCS}^{\rm (cl)}&=& \frac{\Delta^2}{g}+\sum_{j=1}^{N_b} \ve_j -\mu(N-2M) - \sum_{j=0}^{N_b} \sqrt{(\ve_j-\mu)^2+\Delta^2}\label{bcs_cl_gs}\\
\frac{2}{g}&=& \sum_{j=0}^{N_b} \frac{1}{\sqrt{(\ve_j-\mu)^2+\Delta^2}}\label{g_bcs}\\
N-2M&=& \sum_{j=0}^{N_b} \frac{\ve_j-\mu}{\sqrt{(\ve_j-\mu)^2+\Delta^2}}\label{sz_bcs}\; .
\ee
Here $\Delta$ is the superconducting gap, $\mu$ the chemical potential and $H_{\rm BCS}^{\rm (cl)}$ the ground state energy of \refeq{hambcs} in the thermodynamic limit. Eqs.~\refeq{g_bcs} and \refeq{sz_bcs} are equivalent to \refeq{szmf}, \refeq{h0mf}, if Eq.~\refeq{coup_eq} and the following relations hold,
\be
\delta=\Delta/\mu,\qquad \nu=1/\mu,\qquad h_0=1/g\,\label{corr},
\ee
and if furthermore, the sign of the $j=0$-term in Eqs.~\refeq{g_bcs} and \refeq{sz_bcs} is changed. The latter condition reflects the fact that the ground state of the central spin model corresponds to a special single-pair excited state of the BCS-model. This point will be discussed quantitatively in Sec.~\ref{sec:corr}. 

The mechanism of spontaneous symmetry breaking in the classical/mean field approach is completely equivalent in both the BCS and central spin models. In order to see this, we use the pseudospin representation \refeq{ps1}, \refeq{ps2} to write the BCS gap in the pair-excited state corresponding to the ground state of the central spin model as
\be
\Delta=g \sum_{j=1}^{N_b}m_j^\perp.
\ee
Inserting the last of relations \refeq{mj} and substituting Eq.~\refeq{coup_eq} and the first two equations from \refeq{corr}, one re-obtains the gap equation \refeq{g_bcs}. 

An important difference to the BCS-solution consists in the order of magnitude of $h_0=1/g$. To obtain a well-defined energy per particle in the thermodynamic limit, $1/g=\Or(N)$ scales with the number of particles. In the quantum dot, however, the experimental situation corresponds to $h_0=\Or(1)$, thus not scaling with any extensive parameter. It is instructive though to consider the limit of infinite central magnetic field, shown in Fig.~\ref{fig:mag_prof_infinite_field}. Then $\delta\to 0$ and from Eq.~\refeq{h0mf}, $h_0\approx \sum_{j=1}^{N_b} |2(\ve_j-\mu)|^{-1}\gg 1$. Furthermore, Eq.~\refeq{szmf} yields $A_{N/2}<\nu<A_{N/2+1}$, so that Eqs.~\refeq{m0} and \refeq{mj} reproduce the magnetization profile shown in Fig.~\ref{fig:mag_prof_infinite_field}. In this extreme limit, quantum fluctuations are suppressed completely and the classical picture is exact. {Accordingly, the classical mean field approximation is 
generally better justified for the BCS model.}
However, for general fields $h_0=\Or(1)$, apart from the classical contribution discussed in this section, {important quantum fluctuations will occur as well as will be shown in the next section. }

\subsection{Analytical results: Small field limit}
{Eqs.~\refeq{szmf} and \refeq{h0mf} can be solved numerically to determine the parameters $\nu$, $\delta$ from which the magnetization profile Eq.~\refeq{mj} and the two-point-function Eq.~\refeq{sosj_class} are obtained. 
{However, in the physically most relevant limit of small central fields and large particle numbers it is useful to derive approximate analytical expressions} for the one- and two-point correlators. Therefore, we will first calculate the parameters $\delta$, $\nu$ from Eqs.~\refeq{szmf} and \refeq{h0mf} to leading order in $h_0$, before inserting these results into Eqs.~\refeq{mj} 
and \refeq{sosj_class} for the correlation functions.} 

According to Eq.~\refeq{h0mf}, a small central magnetic field corresponds to 
\be
h_0=\frac{N_b\nu}{2\delta^{(1)}}\;,\qquad \delta^{(1)}=\frac{N_b\nu}{2h_0}\label{delta1}\,,
\ee
where the index $\delta^{(1)}$ is the leading term of $\delta$ in a small-field expansion of $\delta$. Since we derived Eqs.~\refeq{szmf} and \refeq{h0mf} for a large number of nuclei, we restrict ourselves to the terms leading in $N_b$ here. Eq.~\refeq{delta1} is consistent with Eq.~\refeq{corr}: Both imply that a diverging pairing strength in the BCS-pairing model leads to a diverging superconducting gap. 

In the same limit, Eq.~\refeq{szmf} leads to 
\be
2 \delta^{(1)} S^z_{\rm tot} = N-\nu N_b x_{-1},
\ee
where we defined the moments $x_\ell$
\be
N_bx_{\ell}:=\sum_{j=1}^{N_b} A_j^{\ell}.\label{moments}
\ee
{The moments with negative (positive) integers $\ell$ are
determined predominately by the smallest (largest) coupling constants.}
  
We consider here a sample which is not macroscopically polarized, i.e.~$S^z_{\rm tot}=\Or(1)$. The case of macroscopic polarization will be dealt with in Sec.~\ref{sec:corr}. Together with Eq.~\refeq{delta1}, we then obtain for the leading term of $\nu$ for small fields, $\nu^{(1)}$: 
\be
\frac{1}{\nu^{(1)}}&=&x_{-1} + \frac{S^z_{\rm tot}}{h_0}\label{nu1}
\ee
Making the same approximations in the expression for the classical ground state energy, Eq.~\refeq{hcl}, and inserting Eqs.~\refeq{delta1} and \refeq{nu1}, we obtain the leading term for small $h_0$
\be
H_{\rm cl}^{(1)}&=&-\frac{N_b x_1}{4} -h_0 \frac{S^z_{\rm tot}}{N} - \frac{h_0^2}{2 N} x_{-1}\label{hclleading}\; .
\ee
For small central fields, this yields the following expressions for the leading terms in a large-$N$-expansion of classical one- and two-point correlation functions in the ground state:  
\be
\vec m_0\cdot \vec m_j&=& -\frac{1}{4} +\frac{1}{2} \frac{h_0^2}{(A_j N)^2}\label{sosjclas}\\
 m_0^z &=& \frac{S^z_{\rm tot}}{N} + \frac{h_0x^{(0)}_{-1}}{N}\label{sozclas}\\
m_j^z&=& - m_0^z+\frac{h_0}{N A_j}\label{sjzclas}\; ,
\ee
where $x^{(0)}_{-1}$ is the leading term in an asymptotic expansion of $x_{-1}$ in the inverse particle number, $x^{(0)}_{-1}:= \int_{0}^1 1/A(x N_b) \d x$, and $A(x N_b)\equiv A_j$ is treated as a continuous function of $x$. 
 
\subsection{Quantitative comparison with numerical results}
\begin{figure}
\setlength{\unitlength}{1cm}
\begin{picture}(1,5)
\put(0.6,4.4){$m_j^z$}
\end{picture}
\psfrag{s}{$j$}
\psfrag{h}{$h_0$}
\includegraphics[width=0.7\textwidth]{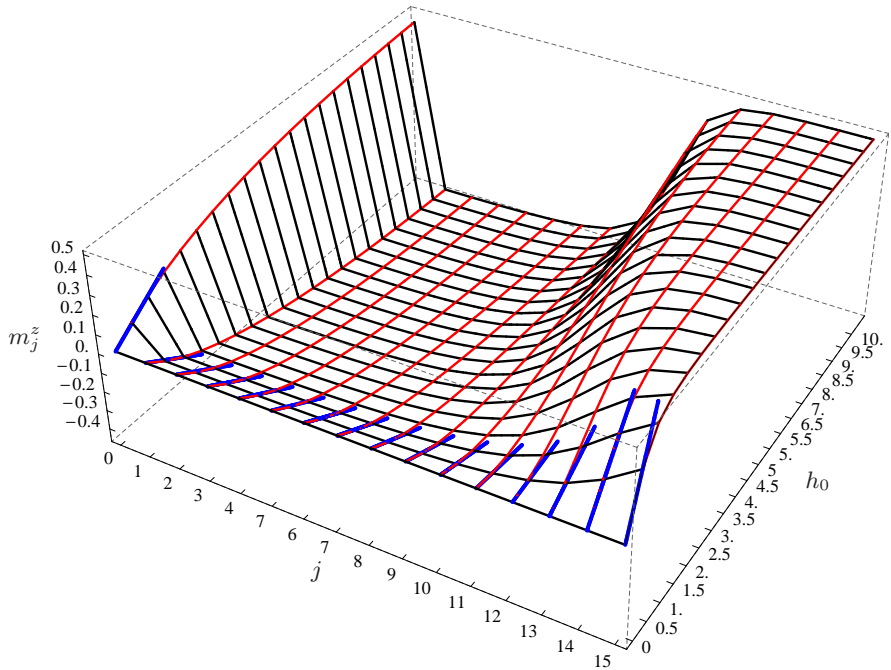}
\caption{(color online) Magnetization profiles as a function of the site $j$ and the central field $h_0$ for $S^z_{\rm tot}=0$ with the couplings \refeq{coup} where $x_1=2$, $B=2$, obtained by inserting the numerical solution of Eqs.~\refeq{h0mf} and \refeq{szmf} for $\delta,\nu$ into Eqs.~\refeq{m0} and \refeq{mj} for $m_j^z$. The short blue lines denote the leading contribution for small fields and large particle number, Eqs.~\refeq{sozclas} and \refeq{sjzclas}.}
\label{fig:classic_cloud}
\end{figure}

We illustrate the classical results in Fig.~\ref{fig:classic_cloud}, where magnetization profiles $m_{j=0,\ldots,N_b}^z$ are shown, after solving Eqs.~\refeq{szmf} and \refeq{h0mf} numerically for $S^z_{\rm tot}=0$ and different $h_0$. The small-field asymptotes from Eqs.~\refeq{sozclas} and \refeq{sjzclas} are depicted as well.    

We now discuss the question to what extent these classical expressions can be identified with the quantum-mechanical expectation values for large particle number and small central field. 

In order to do so, we first compare our results with a complete diagonalization study for a system with $N=16$ sites {as an additional independent check. The coupling constants in this system were chosen according to Eq.~\refeq{coup} with $x_1=2,\,B=2$. In the next section, we will see that the complete diagonalization study also enables us to classify low-lying excited states according to the distribution of the corresponding BA roots, which is not possible a priori.}

\begin{figure}
\psfrag{a}{$A_j$}
\psfrag{j}{$j$}
\psfrag{m}{$\langle S_j^z\rangle_0$}
\includegraphics[width=0.7\textwidth,angle=0]{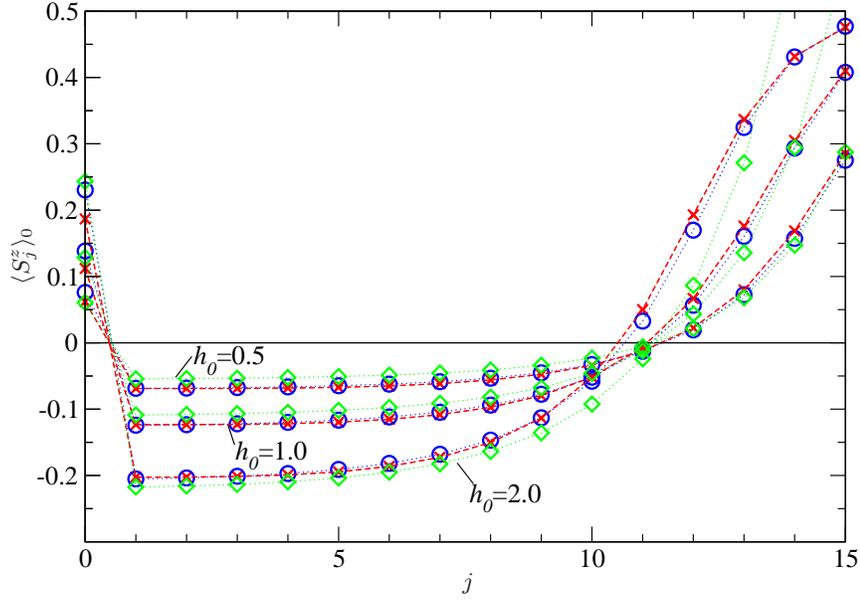}
\caption{{(color online) The local magnetization $\langle S_j^z\rangle_0$ obtained for a $N=16$-spin system with couplings as in Fig.~\ref{fig:mag_prof_infinite_field}. The central field assumes values $h_0=0.5, 1,$ and 2, as indicated in the figure. The 
total polarization is fixed at $S^z_{\rm tot}=0$. Data from complete diagonalization (red crosses) are compared to the small-field expressions Eqs.~\refeq{sozclas}, \refeq{sjzclas} (green diamonds) and the mean-field result Eq.~\refeq{mj} (blue circles), where $\delta, \,\nu$ were obtained by 
numerically solving Eqs.~\refeq{szmf} and \refeq{h0mf}.}}
\label{fig:mag_profile}
\end{figure}

In Fig.~\ref{fig:mag_profile} we compare the diagonalization results with the full classical expressions Eqs.~\refeq{m0} and \refeq{mj} and with the approximate results \refeq{sozclas}, \refeq{sjzclas} for three different values of $h_0$. The small-field expansion \refeq{sjzclas} deviates from the exact data essentially at large distances from the center of the dot, where the more weakly bound spins are located. On the other hand, Eq.~\refeq{mj} with values for $\delta,\,\nu$ obtained by solving Eqs.~\refeq{szmf} and \refeq{h0mf} numerically deviates from the exact solution only by a few percent or less.

\begin{figure}
\psfrag{a}{$A_j$}
\psfrag{j}{$j$}
\psfrag{m}{$\langle \vec S_0\cdot \vec S_j\rangle_0$}
\vskip1cm
\includegraphics[width=0.7\textwidth,angle=0]{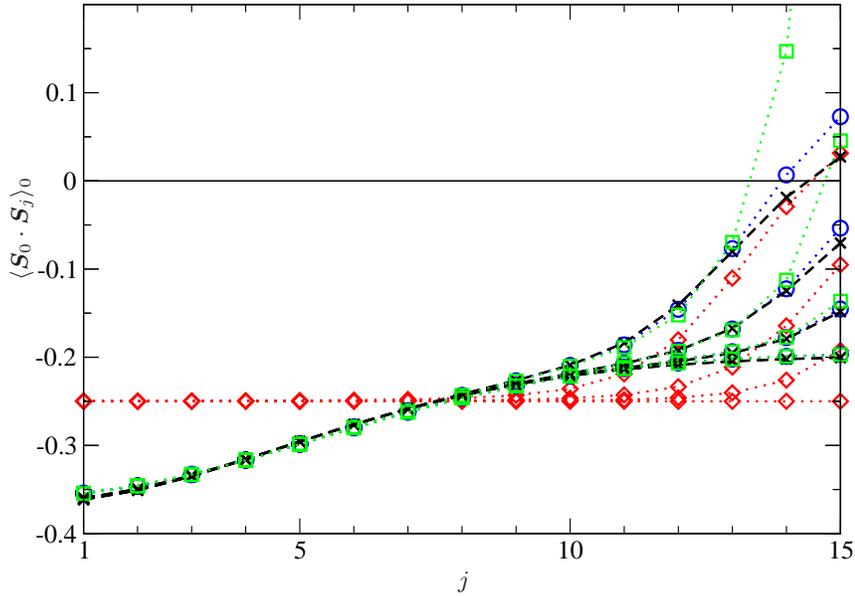}
\caption{(color online) Two point function $\langle \vec S_0\cdot\vec S_j\rangle_0$ in the ground state for the model with the same exchange couplings chosen as in Fig.~\ref{fig:mag_prof_infinite_field} and central field values $h_0=0,0.5,1.0, 2.0$ from below. Black crosses were computed from a complete diagonalization. Green squares were obtained from Eq.~\refeq{sosjsmallh}. Red diamonds stem from the classical solution, Eq.~\refeq{sosj_class}. Blue circles are based on Eq.~\refeq{sosj_class_qm}, where classical and quantum contributions are summed.}
\label{fig:sosj}
\end{figure}

Comparing the classical expression for the two-point function Eq.~\refeq{sosj_class} with the diagonalization results, one notices considerable differences, see Fig.~\ref{fig:sosj}. 
\label{earlier}

\section{Correlation functions from the exact quantum-mechanical solution}
\label{sec:corr}
Whereas in the previous section a classical picture of the central spin model was sketched, this section contains a systematic study of the exact quantum-mechanical solution, where the contribution of quantum fluctuations to the correlation functions will be emphasized. We will first obtain approximate analytical expressions for correlation functions in the regimes of zero and weak central magnetic fields, before recovering the classical picture from the previous section {in the appropriate limit}. 

\subsection{No field}
For vanishing magnetic field, the Hamiltonian \refeq{gaudef} is $SU(2)$-invariant and commutes with all components of the total spin, $\l[H|_{h=0}, \vec S_{\rm tot}\r]=0$. In other words, all states within one spin multiplet, obtained by acting with $S^\pm_{\rm tot}$ on highest weight states, are energetically degenerate. In the expression for the eigenstates \refeq{baes}, application of $S^-_{\rm tot}$ corresponds to choosing $\omega_{k,\nu}=0$. Indeed, for $h_0=0$, it is easy to see that if $\l\{\omega_{1},\ldots,\omega_M\r\}$ is a solution of the coupled set of equations \refeq{bae}, then $\l\{\omega_{1},\ldots,\omega_M,0\r\}$ is a solution as well. Both these solutions are energetically degenerate, according to 
Eq.~\refeq{ev}. This situation is analogous to the Heisenberg chain, where sets of only finite roots encode the highest weight states.\cite{tak99,hag07} Here and in the following, the ground state energy and expectation values in the ground state will be labeled by the subscript $_0$.

\subsubsection{Ground state}
The ground state maximizes $\sum_{k=0}^{M_b}\o_k$. It turns out that the corresponding highest weight state has $M_b=0$, so that only one Bethe number $\omega_0$ has to be determined from $1-\sum_{j=1}^{N_b} \frac{A_j}{\o_0-A_j}=0$. We are interested in the energy levels for large particle numbers. In the ground state, $\omega_0=\Or(N)$, which allows to rewrite the single BA equation in terms of the moments $x_n$ defined in Eq.~\refeq{moments},
\be
1-N_b \sum_{n=1}^\infty \frac{x_n}{\omega_0^n}&=&0\label{omeq}\,.
\ee  
We define $\o_0 =:\wt \o_0 N_b$, such that both $x_n$ and $\wt \o_0$ are
$\Or(1)$. Then, according to (\ref{ev}), the ground state energy reads
\be
E_0=-\frac{N_b}{2}\l(\wt \o_0-\frac{x_1}{2}\r)\label{en0}
\ee
and successive orders of $\wt \o_0$ in an asymptotic expansion for
large particle numbers can be obtained by inverting Eq.~\refeq{omeq}
order by order. {Let $\wt \o_0^{(n)}$ be the expansion of $\wt \o_0$ in
powers of $N_b^{-1}$ up to order $n$, i.e.} $\lim_{N_b\to \infty} \l(\wt \o_0-\wt \o_0^{(n)}\r)N_b^{n+1}=\Or(1)$. For $n=3$, with $d:=(N_b\,x_1)^{-1}$, $y_n:=x_{n+1}/x_1$, we obtain
\be
\frac{\wt \o_0^{(3)}}{x_1}&=&1+y_1 d +(-y_1^2+y_2)d^2+(2y_1^3-3y_1y_2+y_3)d^3\,.\label{serieso}
\ee
This leads us to conjecture that the coefficient of $d^n$ in the expansion of $(\wt\o_0/x_1-1)$ is given by the $n$th coefficient in a Taylor expansion of $n! \ln\l[\phi(d)/x_1\r]$ in the variable $d$, where the generating function is $\phi(d)=\sum_{j=1}^{N_b} A_j \exp\l[d A_j\r]/N_b$. 
 
Before continuing, let us make two comments on Eq.~\refeq{serieso}: To begin with the leading term, when plugged into Eq.~\refeq{en0}, yields the overall ground state energy of the classical model \refeq{hclangles} with no fields, where the central spin is pointing in the direction opposite to the nuclear bath spins. Finite-size corrections, given by the {sub}-leading terms in Eq.~\refeq{serieso} therefore represent quantum effects. Secondly, for the homogeneous model $A_j\equiv A \,\forall j$, all but the first two terms on the right hand side of Eq.~\refeq{serieso} vanish.

{It is now straight-forward to evaluate the moments for a given distribution of the coupling constants $A_j$.
For the particular case of the choice in Eq.~\refeq{coup} it is possible to use the Euler MacLaurin summation formula
to find an expansion of the moments in the parameter $d$.}
Writing $y_1=y_1^{(0)}+d y_1^{(1)}$, we find
\be
y_1^{(0)}&=&2x_1\frac{B\,{\rm Erf}(\sqrt2\,B)}{\sqrt{2\pi}\,{\rm Erf}^2(B)}\\
y_1^{(1)}&=&-\frac{2x_1^2B^2}{\pi {\rm Erf}^2(2)}+4\,B^2\frac{{\rm Erf}(\sqrt2\,B)}{\pi\sqrt2 {\rm Erf}^3(B)}\,.
\ee
Therefore, the coefficient of $d^2$ in Eq.~\refeq{serieso} becomes
\be
\l[-\l(y_1^{(0)}\r)^2+y_2^{(0)}\r]&=&\frac{2x_1^2\,B^2}{\pi {\rm Erf}^3(2)}\l[-{\rm Erf}^2(\sqrt2\,B)/{\rm Erf}(B)+2{\rm Erf}(\sqrt3\,B)/\sqrt3\r].
\ee
As expected, this latter expression tends to zero for $B\to 0$, which is the homogeneous limit in the couplings \refeq{coup}.

From Eq.~\refeq{en0}, one then obtains for the ground state energy
\be
E_0=-\frac{1}{4d} - \frac{y_1^{(0)}}{2} - \frac{d}{2}\l[y_1^{(1)} -\l(y_1^{(0)}\r)^2+y_2^{(0)}\r]\,.\label{e0h0}
\ee
The first term is the classical result, where the central spin is
aligned antiferromagnetically with respect to the bath spins. The second and third terms constitute quantum corrections. 

\begin{figure}
\psfrag{m}{$\langle \vec S_0\cdot \vec S_j\rangle$}
\psfrag{j}{$j$}
\includegraphics[width=0.7\textwidth,angle=0]{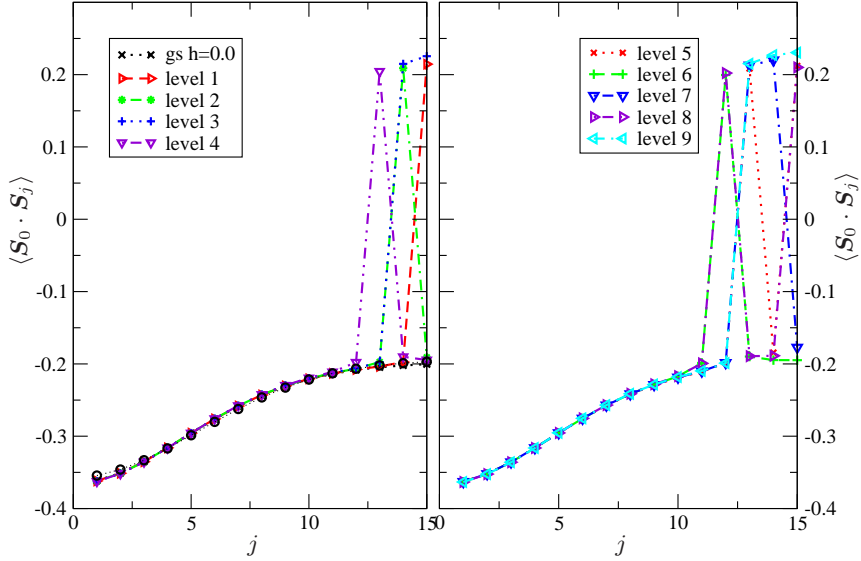}
\caption{(color online) The two-point function $\langle \vec S_0\cdot\vec S_j\rangle$ for zero magnetic field in the ground state (gs) and the lowest ten excited levels (couplings as in Fig.~\ref{fig:mag_profile}), obtained from complete diagonalization. The circles in the left panel are the analytical result Eq.~\refeq{sosjh0} for the ground state.}
\label{fig:sosj_ex}
\end{figure}

In order to calculate the two-point correlation function $\langle \vec S_0 \cdot \vec S_j\rangle_0$ in the ground state, one combines Eq.~\refeq{2point} with Eq.~\refeq{e0h0}. This yields up to order $d$
\be
\langle \vec S_0\cdot \vec S_j\rangle_0=-\frac{1}{4}  +\frac{d}{2} y_1^{(0)}-d A_j \label{sosjh0}.
\ee
Again, the leading contribution reflects the classical picture of
antiferromagnetically aligned spins. For zero central field, quantum fluctuations lead to a
non-trivial dependence on the distance {$j$} between the bath and
the central spins. This is a pure quantum
effect, as can be seen by comparison with
Eq.~\refeq{sosjclas}. Especially, quantum fluctuations decrease
$\langle \vec S_0\cdot \vec S_j\rangle_0$ below the classical result
$-1/4$ if $A_j>y_1^{(0)}/2$, i.e., for the strongly coupled bath
spins. In Fig.~\ref{fig:sosj_ex} the result (\ref{sosjh0}) is compared to complete diagonalization data. In the homogeneous case $A_j=A \forall j$, Eq.~\refeq{sosjh0} reduces to the result found in Refs.~[\onlinecite{ric94,ric96}].

{To calculate} the magnetization profile $\langle S_j^z\rangle_0$, we can employ Eq.~\refeq{sjznu} together 
with Eq.~\refeq{serieso}. Let us first consider the case of a fully polarized bath. 
For $S^z_{\rm tot}=N/2-1$, in leading order this leads to 
\be
\langle S_j^z\rangle_0&=&\frac12-d^2 A_j^2, \qquad j=1,\ldots,N_b\label{sjzh0}\\
\langle S_0^z\rangle_0&=&-\frac12+d y_1^{(0)}\label{s0zh0}.
\ee

\begin{figure}
\psfrag{m}{$\langle S_j^z\rangle$}
\psfrag{j}{$j$}
\vskip1cm
\includegraphics[width=0.7\textwidth]{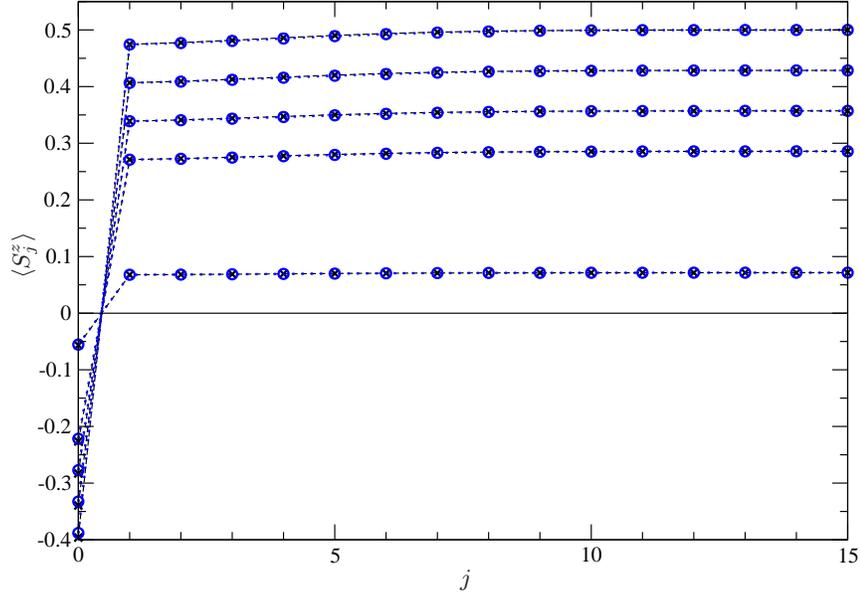}
\caption{(color online) The local magnetization $\langle S_j^z\rangle$ for $N_b=15$ bath spins, with the couplings \refeq{coup} where $x_1=2$, $B=2$. Black crosses stem from a complete diagonalization, blue circles from Eq.~\refeq{sjzh0pol}. For the central spin, $\langle S_0^z\rangle$ was determined such that $S^z_{\rm tot}$ is fixed, which in leading order is given by Eq.~\refeq{s0zh0pol}. From top to bottom, $S^z_{\rm tot}=7,6,5,4,1$. }
\label{fig:szj_h0_pol}
\end{figure}

At smaller values for $S^z_{\rm tot}$, we can still use
Eq.~\refeq{sjznu}, keeping in mind that it has been derived at finite
$h_0$. We thus have to perform the derivative in Eq.~\refeq{sjznu} before
taking the limit $h_0\to 0$. In this limit, the eigenvalues
$\widetilde \Lambda$ in Eq.~\refeq{lambdanu} were  given in Ref.~[\onlinecite{yuz03}], Eq.~(39). Using that result we obtain the expectation values in the respective ground state of each sector $S^z_{\rm tot}$
\be
\langle S_j^z\rangle_{0}&=&\frac{S^z_{\rm tot}}{N-2}\l(1-2d^2 A_j^2\r), \qquad j=1,\ldots,N_b\label{sjzh0pol}\\
\langle S_0^z\rangle_0&=&\frac{S^z_{\rm tot}}{N-2}\l(-1+2d y_1^{(0)}\r)\label{s0zh0pol}. 
\ee
For $S^z_{\rm tot}=N/2-1$, Eqs.~\refeq{sjzh0} and \refeq{s0zh0} are
recovered. In the opposite limit, $S^z_{\rm 0}=0$, the polarization
vanishes, as expected from the $SU(2)$-invariance in this case. In
Fig.~\ref{fig:szj_h0_pol}, we compare the analytical results
\refeq{sjzh0pol}, \refeq{s0zh0pol} with complete diagonalization
data. These illustrate the fact that the magnetization profiles are
different for energetically degenerate states.  

\subsubsection{Excited states}
\label{exh0}
In the classical picture, the lowest excitations at $h_0=0$ above the
{N\'eel-like ground state} are created by {flipping} the spins in the outer region where the exchange with the central spin is weakest. In the exact solution of the quantum mechanical problem \refeq{ev}, \refeq{bae}, excitations can be of two types: 
\begin{itemize}
  \item Spin excitations with a change of $M$, i.e.~the number of roots. 
  \item Particle-hole excitations, where the location of roots is
    changed with respect to the ground state, but the number of roots is kept fixed. 
\end{itemize}
For $h_0=0$, both types of excitations are energetically equivalent: Adding a root $\omega_{k}\neq 0$ is equivalent to moving a root $\omega_{k}=0$ to a finite value. 

Let us consider the excited state where $M_b+1$ roots are different from zero. We focus on low-lying excitations here, so $M_b$ does not scale with the particle number. Thus compared to the ground state for $2S^z_{\rm tot}=N-2(M_b+1)$, there are now $M_b$ additional roots away from the origin. In the set of equations \refeq{bae}, there is one root which scales like the particle number; we denote it by $\o_0$, i.e.~$\o_{0}=\Or(N)$. For the other $M_b$ roots, $\o_{k}=\Or(1)$.  

We define the moments of the additional non-zero roots as $\gamma_n:=\sum_{k=1}^{M_b} \omega^n_{k}$. Performing an expansion analogous to Eq.~\refeq{omeq}, one obtains for the root $\omega_0=\Or(N)$ the equation
\be
1-N_b \sum_{n=1}^\infty \frac{x_n}{\omega_0^n} + 2 \sum_{n=1}^\infty \frac{\gamma_n}{\omega_0^n}=0\label{exex}, 
\ee
which again can be inverted order by order. Including terms of order $\mathcal\Or(1/N_b)$, 
\be
\omega_0=\frac{1}{d}-2\gamma_1+ y_1 +
2dy_1\gamma_1-2d\gamma_2+d(y_2-y_1^2).
\ee
This leads to an expression for the energy in terms of the $\gamma_n$
\be
E_{\rm ex}=-\frac{1}{4d} + {\g_1}- \frac{y_1^{(0)}}{2} - \frac{d}{2}\l[2y_1^{(0)}\gamma_1 + 2\g_2+y_1^{(1)} -\l(y_1^{(0)}\r)^2+y_2^{(0)}\r]\; \label{eex}.
\ee
Let us look at the simplest case, $M_b=1$. The corresponding equation for the additional root $\o_1$ reads
\be
1-\sum_{j=1}^{N_b} \frac{A_j}{A_j-\o_1}+2\frac{\o_1}{\o_0} +2\frac{\o_1^2}{\o_0^2}=0\label{exroot}\; .
\ee
By sketching the lhs of this equation, one sees that $\omega_1$ is located between two couplings. Indeed, for the lowest excitation, we can set $\omega_1=A_{N_b}+\delta_{N_b}$. In leading order, we then obtain $\delta_{N_b}=A_{N_b}/(\sum_{j=1}^{N_b-1}A_j/(A_j-A_{N_b})-1)=\Or(1/N_b)$ and $\delta_{N_b}>0$. One can generalize this result to $\omega_1=A_{\ell}+\delta_{\ell}$, as long as $\delta_\ell=\Or(1/N_b)$ and $\delta_\ell>0$, i.e. for $\ell\gg 1$. Then Eq.~\refeq{sosjh0} is modified according to
\be
\langle \vec S_0\cdot \vec S_j\rangle_{\rm ex} &=& \l\{\begin{array}{ll} -\frac14 + \frac{d}{2} y_1^{(0)} - d A_j,& j\neq \ell\\
\frac14 - \frac{d}{2} y_1^{(0)} + d A_j,& j= \ell\end{array}\r.\label{sosj_ex}.
\ee
This corresponds to the classical picture of spin flips with respect to the ground state at the outer edges of the quantum dot. That result generalizes further to the case of more than one excitation, $M_b>1$. If more than one BA root is present, different root patterns are possible. Let us call the distance $A_{j}-A_{j+1}$ the $j$th coupling interval. We call an interval occupied if one root is located within this interval. 

One type of root configurations consists in only real roots and occupied intervals, with no consecutive occupied intervals. Another type of root configurations involves consecutive occupied intervals. However, depending on the special choice of the coupling constants and the central magnetic field, roots in such a configuration can be driven into the complex plane, thus forming complex conjugate pairs.\cite{ric65,rom04,dom06}

From these observations we conclude that the two-point function $\langle \vec S_0\cdot \vec S_j\rangle_{\rm ex}$ yields significant insight into the underlying root configuration of a low-lying excited state. {\em Vice versa}, if the root configuration for low-lying excitations is known, the corresponding two-point function can be predicted at least qualitatively. This prediction confirms nicely the physical expectation. 

In Fig.~\ref{fig:sosj_ex}, we depict $\langle \vec S_0\cdot \vec S_j\rangle_{\rm ex}$ for the lowest nine excited states for $N=16$ particles with the couplings chosen according to Eq.~\refeq{coup} with $x_1=2$, $B=2$. The data have been obtained from complete diagonalization. The analytical result \refeq{sosjh0} for the ground state is given as well, from which the analytical predictions for excited states are obtained straightforwardly by changing the sign of the corresponding spins, like in Eq.~\refeq{sosj_ex}. 
   
\begin{figure}
\psfrag{a}{$A_{N_b}$}
\psfrag{f}{$A_{N_b-5}$}
\includegraphics[width=0.7\textwidth,angle=0]{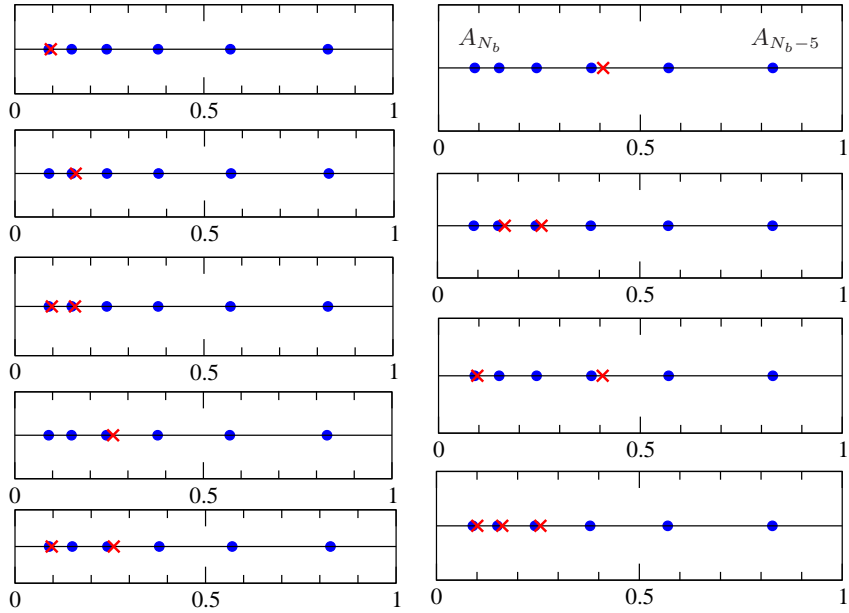}
\caption{(color online) The location of the additional BA numbers which are $\Or(1)$
  (red crosses) for the lowest nine excitations (top left: level 1,
  bottom left: level 5) with $N_b=15$, $x_1=2$, $B=2$ in
  Eq.~\refeq{coup} for the couplings. The six smallest  couplings are shown here (blue dots). Not shown is $\omega_0=\Or(N_b)$.} 
\label{fig:root_config_h0.0}
\end{figure}

It is instructive to consider the corresponding root configurations of those lowest nine excited levels. These are shown in Fig.~\ref{fig:root_config_h0.0} for the highest weight states, i.e.~without roots in the origin. The physical interpretation of the root locations as spin flips with respect to the ground state is revealed when comparing the root pattern level by level with the $j$-dependence of the two-point function. 

This interpretation carries over to the magnetization profile. We show those magnetization profiles corresponding to the lowest nine excited levels in Fig.~\ref{fig:mag_profile_ex_h0.0}. However, as in the ground state, an important difference consists in the degeneracy of $\langle \vec S_0\cdot \vec S_j\rangle_{\rm ex}$ for all states within one multiplet. Whereas the two-point function is independent of the total magnetization $S^z_{\rm tot}$, the local magnetization $\langle S_j^z\rangle_{\rm ex}$ does depend on that quantity. In Fig.~\ref{fig:mag_profile_ex_h0.0}, we only give the magnetization profiles for the highest weight states parametrized by the roots sketched in Fig.~\ref{fig:root_config_h0.0}. By adding additional roots in the origin, i.e.~by lowering $S^z_{\rm tot}$, the two-point function is not altered, but $\langle S_j^z\rangle_{\rm ex}$ is changed by an overall prefactor like in Eqs.~\refeq{sjzh0pol} and \refeq{s0zh0pol}. Namely, proceeding similarly as in the derivation of Eqs.~\refeq{sjzh0pol} and \refeq{s0zh0pol}, one obtains the leading terms of the magnetization profile in low-lying excited states
\be
\langle S_j^z\rangle_{\rm ex}&=&\l\{\begin{array}{cc}
\frac{S^z_{\rm tot}}{N-2n}\l(1-2 d^2 A_j^2\r),& j\neq \ell_1,\ldots,\ell_n\\
-\frac{S^z_{\rm tot}}{N-2n}\l(1-2 d^2 A_j^2\r),& j= 0,\ell_1,\ldots,\ell_n\end{array}\r..\label{mag_ex}
\ee
Here $n\geq 1$ is the number of non-zero roots. {For low-lying states, these are located close to the couplings $A_{\ell_1},\ldots, A_{\ell_n}$, as can be seen from Fig.~\ref{fig:root_config_h0.0}.} All states with the
same $n$ but different $S^z_{\rm tot}$ are energetically degenerate,
i.e.~have identical two-point functions $\langle \vec S_0\cdot \vec
S_j\rangle_{\rm ex}$, but different magnetization profiles $\langle
S_j^z\rangle_{\rm ex}$.  For the highest weight states we have 
$S^z_{\rm tot}=N/2-n$, {which is also the case for the root configurations depicted in
Fig.~\ref{fig:root_config_h0.0}. According to Eq.~\refeq{mag_ex} the magnetization profile therefore {can be read off from the number and location of Bethe roots.}
This is confirmed by
complete diagonalization data shown in
Fig.~\ref{fig:mag_profile_ex_h0.0}. Deviations from Eq.~\refeq{mag_ex}
are due to interactions between the excitations, which were  neglected in the derivation of Eq.~\refeq{mag_ex}.

\begin{figure}
\psfrag{j}{$j$}
\psfrag{m}{$\langle S_j^z\rangle$}
\includegraphics[width=0.7\textwidth]{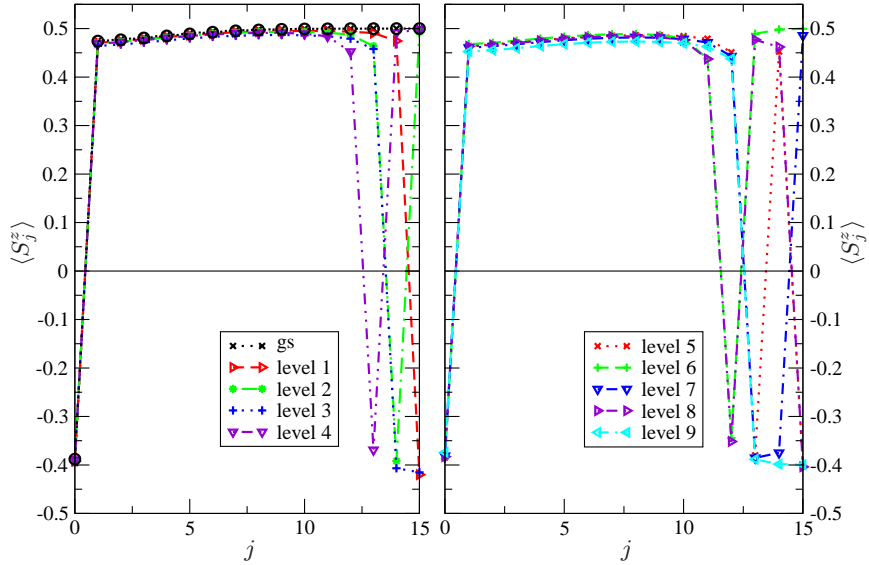}
\caption{(color online) The magnetization profile $\langle S_j^z\rangle_{\rm ex}$ in the ground state (gs) and the lowest nine excited levels corresponding to the root configurations of Fig.~\ref{fig:root_config_h0.0}, obtained from complete diagonalization. The circles for the ground state in the left panel are the analytical results from Eq.~\refeq{sjzh0pol}. The analogous results for excited states agree with Eq.~\refeq{mag_ex} to the same order.}
\label{fig:mag_profile_ex_h0.0}
\end{figure}

\subsection{Weak field}
A finite magnetic field couples to both the electronic and nuclear
spins. As described below Eq.~\refeq{ham}, the coupling to the nuclear
spins is trivial and can be {accounted for at the end of the
  calculation}. Let us thus first focus on $h_0\gtrsim 0$, 
$h_{\rm t}=0$. As stated in the previous section, for $h_0=0$, the
eigenvalues of a given multiplet are degenerate, which in the root
pattern is encoded by roots in the origin. For $h_0\neq 0$, this
degeneracy is lifted due to the broken $SU(2)$ invariance. Thus it is reasonable to assume that the zero roots are driven
{away from} the origin by a finite magnetic field $h_0$. In the
weak field limit, this is confirmed by the large-$g$-expansion of the
Bethe roots\cite{yuz03} of the BCS-Hamiltonian \refeq{hambcs}, which are related
to the Bethe roots of the central spin model via Eq.~\refeq{ekok}. In
this section, we will derive the {energy} eigenvalues and
expressions for the screening cloud and magnetization profile for
small but finite central field. 

\subsubsection{Ground state}
Starting from Eq.~\refeq{bae}, we include a finite field $h_0$ in Eq.~\refeq{exex}. We then arrive at the following equations for $\o_{k=0,\ldots,N_b}$ 
\be
1-N_b\sum_{n=1}^\infty \frac{x_n}{\omega^n_0} +2 \sum_{n=1}^\infty \frac{\gamma_n}{\omega^n_0}+\frac{2 h_0}{\omega_0}&=&0.\label{o0h}\\
1+\sum_{j=1}^{N_b} \frac{A_j}{A_j-\o_k} -2 \sum_{n=0}^\infty \l(\frac{\o_k}{\o_0}\r)^n-2\sum_{k'=1\atop k\neq k'}^{M_b}\frac{\omega_{k'}}{\omega_{k'}-\omega_k}+\frac{2h_0}{\o_k}&=&0\label{okh}.
\ee
Thus the only effect of $h_0$ in Eq.~\refeq{o0h} compared to Eq.~\refeq{exex}
is to add a term $-2h_0$ to $N_b x_1$, i.e.~up to order $\mathcal\Or(1/N_b)$ we have,} 
\be
\omega_0=N_b x_1 -2h_0 -2 \gamma_1(1-dy_1) + y_1(1+2h_0d)-2 \gamma_2 d+
d(y_2-y_1^2).\label{om0smallh0}
\ee
 One then obtains for the ground state energy an expression which still involves the $M_b$ non-zero roots
\be
E_0=-\frac{1}{4 d}+ \frac{h_0}{2} + \frac{\gamma_1}{2} - \frac{y_1}{2} - 
\frac{d}{2}\l[y_2-y_1^2+h_0 y_1 +y_1\gamma_1 -\gamma_2\r]\label{ensmallh0}.
\ee
Let us now multiply Eq.~\refeq{okh} by $\omega_k$ and sum all terms $k=1,\ldots,M_b$. We assume that $h_0$ is sufficiently small so that max$\{|\o_k|\}<A_{N_b}$ and find
\be
\g_1+2\sum_{j=2}^\infty \frac{\gamma_j}{\omega_0^{j-1}}=N_b \sum_{j=1}^\infty x_{-j}\gamma_{j+1}\label{ans},
\ee
with $x_0\equiv 1$. Here, we aim at calculating the energy up to $\Or(h_0^2)$. In analogy to Ref.~[\onlinecite{yuz03}], we therefore make the Ansatz
\be
\gamma_1&=&c_1 h_0 + c_2 h_0^2 + \Or(h_0^3)\label{gamma1}\\
\gamma_2&=& d_1 h_0^2+ \Or(h_0^3)\label{gamma2}.
\ee
Then, including terms $\Or(h_0^2)$, the coefficients $c_{1,2}$ are found by inserting that Ansatz into Eq.~\refeq{ans},
\be
c_1&=&-\frac{2M_b}{N_b-1}\label{c1}\\
c_2&=&\frac{1}{N_b-1}\l(2d-x_{-1}\r) d_1=-\frac{x_{-1}}{N_b-1} d_1+\Or(d^2)\label{c2},
\ee
where we only keep the leading finite-size terms. 

An additional equation is thus needed to determine $d_1$. This is obtained by {adapting} the techniques used in 
Ref.~[\onlinecite{yuz03}] to our problem. We then find that in leading order in $h_0$, the roots $\omega_{k=1,\ldots,M_b}$ are related to the zeros of associated Legendre polynomials,
\be
L^{-N_b}_{M_b}\l( \frac{2h_0}{\o_k}\r)=0.
\ee
This is a polynomial of degree $M_b$, i.e. $L^{-N_b}_{M_b}(x)\equiv c \prod_{k=1}^{M_b}(x-2 h_0/\o_k)$, where the constant $c$ is determined by the asymptotes. Consequently, the logarithmic derivative at $x=0$ is
\be
\l.\l(\ln L^{-N_b}_{M_b}(x)\r)'\r|_{x=0}=-\frac{\gamma_1}{2h_0}\label{lnprime}\;.
\ee
On the other hand, $\l.\l(\ln L^{-N_b}_{M_b}(x)\r)'\r|_{x=0}=M_b/(N_b-1)$, which in combination with Eq.~\refeq{lnprime} confirms Eq.~\refeq{c1}. Analogously, the second logarithmic derivative $\l.\l(\ln L^{-N_b}_{M_b}(x)\r)''\r|_{x=0}=M_b(M_b+1-N_b)/\l((N_b-2)(N_b-1)^2\r)$, which leads to 
\be
\frac{M_b(M_b+1-N_b)}{(N_b-2)(N_b-1)^2}=-\frac{\gamma_2}{(2 h_0)^2}\,.
\ee
Combining this equation with Eq.~\refeq{gamma2}, one finds in terms of $S^z_{\rm tot}$
\be
d_1&=& \frac{(N_b-1)^2-4\l(S^z_{\rm tot}\r)^2}{(N_b-2)(N_b-1)^2}\label{d1}\\
c_2&=&-x_{-1}\frac{(N_b-1)^2-4\l(S^z_{\rm tot}\r)^2}{(N_b-2)(N_b-1)^3}N_b\label{cc2}\,,
\ee
where the latter relation follows from Eq.~\refeq{c2}. Then the ground state energy reads
\be
E_0&=&-\frac{1}{4d}-\frac{y_1}{2} + \frac{S^z_{\rm tot}}{N_b-1}h_0 -2 \frac{S^z_{\rm tot}}{N_b-1}d y_1h_0 +\frac{c_2}{2}h_0^2-\frac{d}{2}(y_2-y_1^2)+\Or(d^2)\nn\\
&=& -\frac{1}{4d} - \frac{y_1^{(0)}}{2}-\frac{d}{2}\l[y_1^{(1)}-\l(y_1^{(0)}\r)^2+y_2^{(0)}\r]+ s^z_{\rm tot}h_0(1-2d y_1) -\frac{1-4 \l(s^z_{\rm tot}\r)^2}{2N_b}x_{-1}^{(0)}h_0^2+\Or(d^2)\label{ensmallh},
\ee
{where we have defined the total magnetization density $s^z_{\rm tot}:=S^z_{\rm tot}/(N_b-1)$}. In the last equation, the leading orders in a finite-size and small-$h_0$ expansion are given. For $S^z_{\rm tot}=0$, the central magnetic field does not enter linearly, but due to second-order spin-exchange processes only quadratically. 

\begin{figure}
\includegraphics[width=0.7\textwidth]{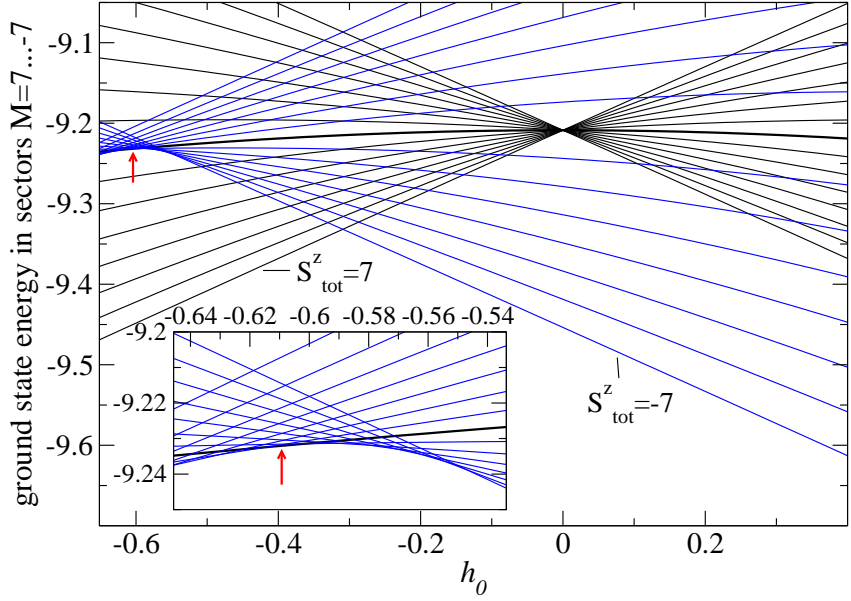}
\caption{(color online) 
Lowest levels in the sectors $M=7\ldots-7$ for $N_b=15$ and couplings as in Eq.~\refeq{coup} with $x_1=B=2$, obtained from a numerical solution of Eq.~\refeq{bae}. Black lines: $h_{\rm t}=0$ (the fat black line is the energy for $S^z_{\rm tot}=0$); blue lines: $h_{\rm t}=0.035$. The inset is a zoom into the region around the caustic. The red arrows mark $h_0$ values such that the state with $S^z_{\rm tot}=0$ is the ground state.}
\label{fig:caustic}
\end{figure}

It is interesting to note that when one adds the additional total
field term $-h_{\rm t}S_{\rm tot}^z$ from Eq.~\refeq{ham} to
Eq.~\refeq{ensmallh},  the lowest levels $E_0(S^z_{\rm tot},h_0)$
{display a pattern which is strongly reminiscent of light rays
  forming a caustic in optics}. This is visualized in Fig.~\ref{fig:caustic}. {Thus}  for a finite total magnetic field $h_{\rm t}$, a small but finite range of values for $h_0$ exists where the state with $S^z_{\rm tot}=0$ is the ground state. Or, coming back to the original Hamiltonian \refeq{gaudef}, this means that for any finite ratio $g_{\rm n}/g_{\rm e}$, one can adjust the field such that the ground state has a given magnetization $S^z_{\rm tot}$. For the example shown in Fig.~\ref{fig:caustic}, the ground state has zero magnetization for $h_{\rm t}=0.035$, $h_0\approx-0.61$, which corresponds to $g_{\rm n}/g_{\rm e}\approx 0.054$ with $h g_{\rm n}=h_{\rm t}$.  

Analytically, the relation between $h_0$ and $h_{\rm t}$ for a given $S^z_{\rm tot}$ is found from $h_{\rm t}=-\partial_{S^z_{\rm tot}} E_0$, with $E_0$ given in Eq.~\refeq{ensmallh}. This leads to
\be
-h_{\rm t}=\frac{h_0}{N_b-1}\l(1-2 d y_1\r) + \frac{4 s^z_{\rm tot} x_{-1}N_b}{N_b-2}\l(\frac{h_0}{N_b-1}\r)^2\,,
\ee
up to higher order corrections. For $S^z_{\rm tot}=0$, this is inverted to 
\be
-h_0=(N_b-1)(1+2 d y_1) h_{\rm t}=\l(N_b+\frac{2x_2^{(0)}}{x_1^2}-1 \r)h_{\rm t}+ \Or(d)\label{h0ht},
\ee
which for the numerical values chosen in Fig.~\ref{fig:caustic} yields
$h_0\approx -0.60$, in good agreement with the exact numerical data
from the Bethe Ansatz. From the classical Hamiltonian
\refeq{hclleading}, only the leading contribution in the particle number in Eq.~\refeq{h0ht} is recovered. From Eq.~\refeq{h0ht}, the ratio of $g$-factors in the $S^z_{\rm tot}=0$-sector is deduced (setting $h g_{\rm n}=h_{\rm t}$), 
\be
\frac{g_e}{g_n} = N_b +\frac{2x_2^{(0)}}{x_1^2}+ \Or(d)\label{gfactors}.
\ee
This means that when the ratio of electronic to nuclear $g$-factors
equals the number of nuclear bath spins, then an overall  magnetic field drives the system into the non-degenerate $S^z_{\rm tot}=0$ state. Since this ratio is of the order $\Or(10^3)$, it is very realistic to probe this regime in an experimental setup. 

Let us now consider correlation functions. From Eq.~\refeq{2point}, we obtain for large particle numbers
\be
\langle \vec S_0\cdot \vec S_j\rangle_0 &=& -\frac14 +\frac{d}{2} y_1 -d A_j-\frac{d^2}{2}\l(-2y_2+3y_1^2+3 A_j^2-4y_1A_j\r)+\frac{c_2^{(0)}}{2}\,\frac{h_0^2}{A_j^2}+\Or(d^3)\label{sosjsmallh}.
\ee
with $c_2^{(0)}=(1-4\l(s^z_{\rm tot}\r)^2)/N_b^2$. 
If one is interested in the coefficients of the asymptotic $1/N_b$-expansion one should again apply the Euler-MacLaurin formula as in Eq.~\refeq{e0h0}. We do not want to dwell into these technical but straightforward details here but rather compare the analytical prediction with exact results from complete diagonalization. Such a comparison is shown in Fig.~\ref{fig:sosj} in Sec.~\ref{earlier}.

As expected, the approximation \refeq{sosjsmallh} is reliable quantitatively only for small fields, {according to 
our weak-field assumption} max$\{|\o_k|\}<A_{N_b}$. 
From Eq.~\refeq{gamma1}, one estimates $|\o_k|\approx 2h_0/N_b$, which means that Eqs.~\refeq{ensmallh} 
and \refeq{sosjsmallh} are valid for $|h_0|\lesssim A_{N_b} N_b/2$. For our choice of parameters $N_b=15$, $x_1=B=2$
in the complete diagonalization of Eq.~\refeq{coup}, this means $|h_0|\lesssim  0.7$. But even for larger values of $h_0$, Eq.~\refeq{sosjsmallh} is qualitatively correct: A finite central magnetic field leads to an enhanced {\em ferromagnetic} correlation between the central spin and the rather loosely bound bath spins at larger distances from the center of the dot, and to an enhanced {\em antiferromagnetic} correlation between the central spin and the bath spins closer to the center of the dot, {which is also consistent with the classical magnetization profile in Fig.~\ref{fig:classic_cloud}.}

We also want to compare Eq.~\refeq{sosjsmallh} with
Eq.~\refeq{sosjclas}, obtained within the classical picture for
$S^z_{\rm tot}=0$ for $h_0\gtrsim 0$. In that approximation, the equations differ from each other by
field-independent terms proportional to $d$. This is understandable: We have
seen in Eq.~\refeq{sosjh0} that these terms constitute finite-size corrections
which stem from quantum fluctuations and are thus not present within the
classical approach. These lead to an increase of the amplitude of two-point
functions. Especially, for the stronger couplings, values smaller than the
classical bound $-1/4$ are reached, a clear sign of entanglement and
non-commutativity of the quantum spin operators. 

In order to determine quantum fluctuations to $\langle S_j^z\rangle$ in the
small-field limit, one has to solve the set of equations~\refeq{auxeq} for the $E_j$ in order
to determine the eigenvalue in Eq.~\refeq{lambdanu}. From this, the local
magnetization is obtained via Eq.~\refeq{sjznu}. The small-field expansion of the set of Eqs.~\refeq{auxeq} has been studied in detail in
Ref.~[\onlinecite{yuz03}]. From that work, it follows that for the ground state of
the central spin model at fixed $N_b$, $M_b$, 
\be
\sum_{k=1}^{N_b} E_k= E_{\rm gr}(M_b, N_b-1,x_{-p}-2 E_0^p/N_b)
\ee
with
\be
E_{\rm gr}(M_b,N_b-1,x_{-p})&=&-\frac{M_b(N_b-M_b)}{2 h_0} +
\frac{M_b\,N_b x_{-1}}{N_b-1}\nn\\
& &-2 h_0(x_{-2}-x_{-1}^2)\frac{N_b
  M_b(N_b-1-M_b)}{(N_b-1)^2 (N_b-2)}+\Or(h_0^2).
\ee
From Eq.~\refeq{om0smallh0} one computes 
\be
E_0=d(1+2dh_0 +2d\gamma_1 + d y_1) + \Or(h_0^2,d^3),
\ee
and $\gamma_1$ follows from combining Eqs.~\refeq{gamma1} and \refeq{c1}. In
linear order in $h_0$ and including orders $\Or(d^3)$, one then obtains the magnetization profile
\be
\langle S_j^z\rangle& =& s^z_{\rm tot}(1-2 A_j^2 d^2-16 h_0 A_j^2 d^3 s^z_{\rm
  tot}+ A_j^2 d^3(3 y_1-2 A_j))\nn\\
& &+ \l(A_j^{-1}- x_{-1}(1-2d^2 A_j^2)+2 d^2 x_1(1-2 d^2
A_j^2)\r)\frac{1-4(s^z_{\rm tot})^2}{N_b-2} h_0\label{sjzsmallh0fin}
\ee
with $s^z_{\rm tot}:=S^z_{\rm tot}/(N_b-1)$. The polarization of the central
spin, $\langle S_0^z\rangle$, is fixed by the sum rule $S^z_{\rm
  tot}=\sum_{j=0}^{N_b} \langle S_j^z\rangle$. Comparing
Eq.~\refeq{sjzsmallh0fin} with Eqs.~\refeq{sozclas} and \refeq{sjzclas}, one
again recognizes the effect of quantum fluctuations which are now sub-leading with respect to the classical contributions. As expected, these reduce
the amplitude of the magnetization profile, signaling the effects of
entanglement. 

\subsubsection{Excited states}
In Sec.~\ref{exh0}, we found excitations for $h_0=0$. We can proceed similarly for $h_0\gtrsim 0$. The expression 
\refeq{ensmallh0} for the energy is still valid, but the $\gamma_{1,2}$ are different now. Let us first consider single-particle excitations, parametrized by a single root $\omega_1$ located on the real axis between $A_{j+1}$ and $A_j$. Instead of Eq.~\refeq{okh} for $k=1,\ldots,M_b$, the corresponding set of equations now reads
\be
1+\sum_{j=1}^{N_b}\frac{A_j}{A_j-\o_1}- 2\sum_{j=0}^\infty \l(\frac{\o_1}{\o_0}\r)^j-2\sum_{j=
1}^\infty \frac{\gamma_j}{\o_1^j}+ \frac{2h_0}{\o_1}&=&0\\
1+ N_b\sum_{j=0}^\infty \o_k^j x_{-j}-2 \sum_{j=0}^\infty \l(\frac{\o_k}{\o_0}\r)^j-2\sum_{j=0}^\infty \l(\frac{\o_k}{\o_1}\r)^j -2 \sum_{k'=2}^{M_b} \frac{\o_{k'}}{\o_{k'}-\o_k}+\frac{2h_0}{\o_k}&=&0,\;\\
& &\hfill k=2,\ldots,M_b.\nn
\ee
By multiplying the latter equation with $\o_k$ and taking the sum $k=2,\ldots,M_b$, one arrives at an equation similar to \refeq{ans}, with $M_b\to M_b-1$, $N_b\to N_b-2$ and $x_{-n}\to x_{-n} - 2/(N_b\o_1^n)$, $n=0,1,\ldots$. Thus the coefficients in Eqs.~\refeq{gamma1} and \refeq{gamma2} are now
\be
c^{\rm ex}_1&=&-2\frac{M_b-1}{N_b-3}\\
c_2^{\rm ex}&=&\frac{1}{N_b-3}\l(2d+\frac{2}{\omega_1}-N_b x_{-1}\r)d_1^{\rm ex}
\ee
with {$d_1^{\rm ex}=d_1(N_b\to N_b-2,S^z_{\rm tot}\to S^z_{\rm tot})$ }
as defined in Eq.~\refeq{d1}. 

\begin{figure}
\psfrag{j}{$j$}
\psfrag{m}{$\langle \vec S_0\cdot \vec S_j\rangle$}
\includegraphics[width=0.7\textwidth]{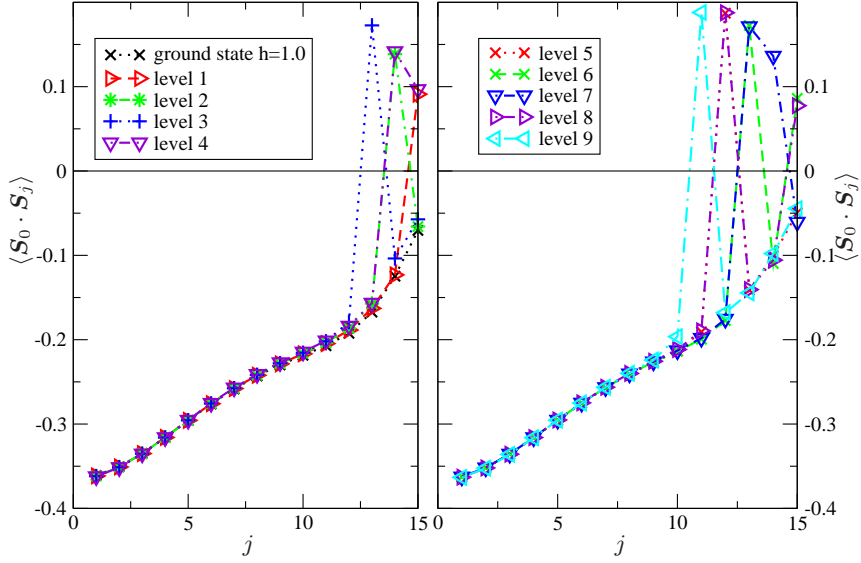}
\caption{(color online) The two-point function $\langle \vec S_0\cdot \vec S_j\rangle_{\rm
  ex}$  for $h_0=1.0$ for the ground state and the
  lowest nine excited states, obtained from complete diagonalization with
  $N_b=15$, and the couplings according to Eq.~\refeq{coup} {with $x_1=B=2$.}} 
\label{fig:sosjh1ex}
\end{figure}

\begin{figure}
\psfrag{j}{$j$}
\psfrag{m}{$\langle S_j^z\rangle$}
\vskip1cm
\includegraphics[width=0.7\textwidth]{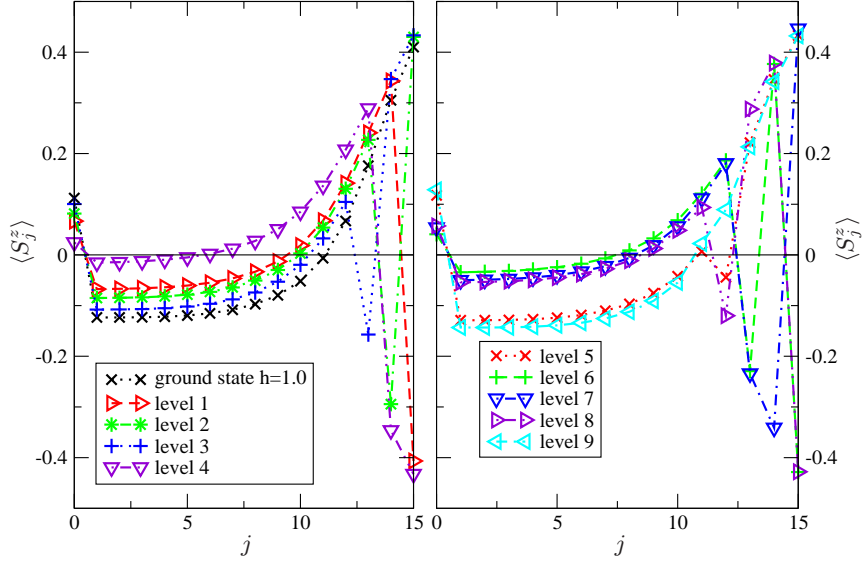}
\caption{(color online) The magnetization profile for $h_0=1.0$ for the ground state and the
  lowest nine excited states, obtained from complete diagonalization with
  $N_b=15$, and the couplings according to Eq.~\refeq{coup} {with $x_1=B=2$.}} 
\label{fig:sjzh1ex}
\end{figure}

\begin{figure}
\psfrag{a}{$A_{N_b}$}
\psfrag{f}{$A_{N_b-5}$}
\vskip1cm
\includegraphics[width=0.7\textwidth]{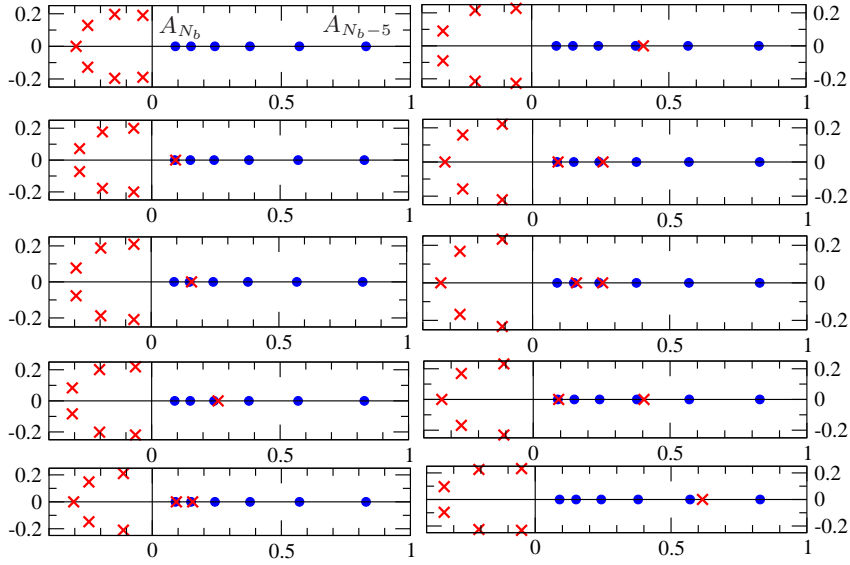}
\caption{(color online) The location of those BA numbers which are $\Or(1)$ (red crosses) in the complex plane for the ground state and the lowest nine excitations in the sector with $S^z_{\rm tot}=0$ (top left: ground state, bottom left: level 4) with $N_b=15$, $x_1=2$, $B=2$ in Eq.~\refeq{coup} for the couplings. The smallest six couplings are shown here (blue dots). Not shown is $\omega_0=\Or(N_b)$.} 
\label{fig:h1exroots}
\end{figure}

A small field does not change the root pattern of the lowest excited states
qualitatively. Very similar to the discussion after Eq.~\refeq{exroot}, one
can still make the Ansatz $\o_\ell=A_\ell+\delta_\ell$ for single particle
excitations. Then for small $h_0$ and $\ell\gg 1$, one again finds that
$\delta_\ell=\Or(1/N_b)$. This picture carries
  over to multiparticle
excitations, except that a finite field $h_0$ can lead to complex 
conjugate pairs of roots.\cite{ric65,rom04,dom06} 
Thus the low-energy excitations are still given by approximately
independent spin flips of the outer bath spins. This is best seen when
comparing the correlation functions $\langle \vec S_0\cdot \vec S_j\rangle_{\rm ex}$,
$\langle S_0^z\rangle_{\rm ex}$ with the root patterns corresponding to the
excited states. In Fig.~\ref{fig:sosjh1ex}, the two-point function $\langle
\vec S_0\cdot \vec S_j\rangle_{\rm ex}$ is shown for the ground state and the lowest
nine levels with a central field $h_0=1$ in the sector $S^z_{\rm tot}=0$. The
corresponding magnetization profile, $\langle S_j^z\rangle_{\rm ex}$ is
sketched in Fig.~\ref{fig:sjzh1ex}, and Fig.~\ref{fig:h1exroots} shows the
underlying root patterns.

Although qualitatively, the results are similar to those shown in
Figs.~\ref{fig:mag_profile} and \ref{fig:root_config_h0.0}, there
are two important differences. Firstly, the degeneracy between states within
one multiplet is lifted, so that both the one- and two-point functions depend on the
total magnetization (in
Figs.~\ref{fig:sosjh1ex} and \ref{fig:sjzh1ex}, we have chosen $S^z_{\rm
  tot}=0$). Secondly, the ordering of root
configurations according to their energies is different. For example, the third
excited level for $h_0=0$ is given by a two-particle excitation (two flipped
spins), as shown in Fig.~\ref{fig:root_config_h0.0}, whereas for $h_0=1.0$, such a
configuration yields the fourth excited level, cf.~Fig.~\ref{fig:h1exroots}. 

\subsection{From the exact solution to the classical picture}
\label{bcs}
\begin{figure}
\psfrag{g}{Re$[\omega_k]$}
\psfrag{h}{Im$[\omega_k]$}
\psfrag{j}{Re$[E_k]$}
\psfrag{i}{Im$[E_k]$}
\psfrag{e}{$\omega_0$}
\psfrag{f}{$E_0$}
\includegraphics[width=0.7\textwidth,angle=0]{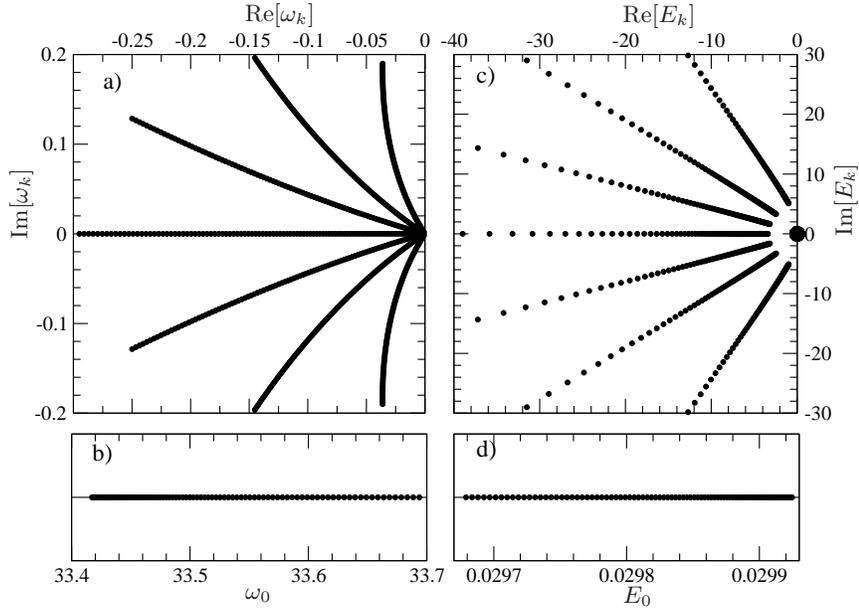}
\caption{a) Bethe roots $\omega_{k}=\Or(1)$, $k=1,\ldots,7$ for $0<h_0\leq 1.0$. The field drives the roots away from the origin; the real root $\o_0$, shown in b), moves towards the origin. c) The inverse numbers $E_k\equiv 1/\omega_k$, including $k=0$ (on the real axis close to the origin, d) is a zoom showing $E_0$ only). Here, the field makes the roots move towards the origin, except for $E_0$, which is shifted to larger real values. }
\label{fig:moving_roots} 
\end{figure}

In this section, we want to make contact with the classical picture presented in section
\ref{tl}, starting from the exact solution for large $N$ and small polarization,
i.e.~$M=\Or(N)$, and finite central field. In this situation, the question
arises whether the Bethe roots form a dense distribution in the complex plane,
which would permit a continuum description. In Fig.~\ref{fig:moving_roots}, we
show both the roots $\omega_k$ and the inverse numbers $E_k=1/\omega_k$ for $N=16,\,M=8$, i.e.~$S^z_{\rm tot}=0$, parametrized by $h_0$. One can show \cite{ric77} that for $g^{-1}=\Or(N)$, the distribution of the $E_k$ can be described by a cut in the complex plane in the thermodynamic limit $M,N\to\infty$, $M/N$ fixed. However, for the central spin model, we are interested in $g^{-1}\equiv h_0=\Or(1)$. Fig.~\ref{fig:moving_roots} suggests that such a continuum description still is possible in this case. To see this, we first review Richardson's \cite{ric77} line of arguments for $g^{-1}\equiv \Or(N)$. 

Consider the function
\be
F(z)=\sum_{k=0}^{M_b}\frac{1}{z-E_k}-\frac12 \sum_{j=0}^{N_b}\frac{1}{z-\ve_j}-g^{-1}\label{deffz},
\ee
where $\ve_j, \,E_k$ are related to $A_j,\,\omega_k$ according to
Eqs.~\refeq{coup_eq} and \refeq{ekok} and $\ve_0=0^-$. Then for
$g^{-1}=\Or(N)$, the function $F$ is expanded as $F=F_0+F_1+\ldots$,
where $F_\nu=\Or\l(N^{-\nu+1}\r)$. Such an expansion is justified
rigourously by showing that $F$ obeys a differential equation which
can be solved order by order. We first consider the case where in the
leading order, {\em all} $E_k$ merge to form a cut in the complex
plane along an arc which is symmetric with respect to the real axis. The endpoints of the arc, $a$ and $a^*$, are parametrized by two real quantities, $a=\mu+\rmi \Delta$, where $\mu$ and $\Delta$ are the chemical potential and the superconducting gap of the BCS model \refeq{hambcs}. This situation corresponds to the ground state of the BCS model. Then 
\be
F_0(z)&=&-\sum_{j=0}^{N_b}\frac{\sqrt{(z-\mu)^2+\Delta^2}}{2(z-\ve_j)\sqrt{(\ve_j-\mu)^2+\Delta^2}}\label{f0z}.
\ee
The ground state of the central spin model corresponds to a particle-hole excited state of the BCS-model, where one root $E_0$ is taken away from the arc and instead is located on the positive real axis, close to $\ve_0=0^-$, namely $E_0=1/\omega_0=\Or(1/N)$. This is shown in the right panel of Fig.~\ref{fig:moving_roots}. Let us thus define $F'(z)=\ov F'(z)+1/(z-E_0)$, where $\ov F'(z)$ contains the roots on the arc. Taking only one root away from the arc does not modify the arc in leading order,\cite{ric77} such that $\ov F_0'(z)=F_0(z)$. This means that in leading order, the roots on the arc are decoupled from $E_0$. 

Let us now focus on the ground state of the central spin model with $h_0=\Or(1)$. If we assume that the roots $E_{k=1,\ldots,M_b}$ in the ground state of the central spin model are still described by an arc in the complex plane for large particle number, we have $\gamma_1=-\ov F'(0)+\frac{1}{2d}+\frac{A_0}{2}-h_0$. This implies for the leading order $\gamma_1^{(0)}$ 
\be
\gamma_1^{(0)}&=&-F_0(0)+\frac{1}{2d}+\frac{A_0}{2}-h_0\\
&=& -\sum_{j=1}^{N_b}\frac{\sqrt{1+\delta^2} A_j^2}{2\sqrt{(\nu-A_j)^2+(A_j \,\delta)^2}}+\frac{1}{2d}-h_0 \label{gamma10},
\ee
where we have used the correspondence between BCS- and central spin parameters in Eqs.~\refeq{coup_eq} and \refeq{corr}. Note that $A_0=1/\ve_0$ drops out in the first line. 

The two parameters $\delta, \mu$ are now determined by the asymptotes of $F_0(z)$, 
\be
h_0&=&-\lim_{z\to \infty} F_0(z)=\frac12 \sum_{j=1}^{N_b}\frac{\nu A_j}{\sqrt{(\nu-A_j)^2+(A_j\,\delta)^2}}-\frac{\nu}{2\sqrt{1+\delta^2}}\\
2 S^z_{\rm tot}&=&-\lim_{z\to\infty}z\l[F_0(z)+h_0\r]=\frac{1}{\sqrt{1+\delta^2}}+\sum_{j=1}^{N_b}\frac{\nu-A_j}{\sqrt{(\nu-A_j)^2+(A_j\,\delta)^2}}\label{sztotqm},
\ee
where we have set $\varepsilon_0=0^-$. 
The first of these equations coincides with Eq.~\refeq{h0mf}. The second equation \refeq{sztotqm} is identical to Eq.~\refeq{szmf}. 

We have verified numerically that $\gamma_1^{(0)}+h_0=\Or(1/N_b)$ for fields $h_0=\Or(1)$. More generally, for $S^z_{\rm tot}=\Or(1)$, the quantity $\gamma_1^{(0)}+h_0$ is of the order $\Or(h_0^2/N_b)$, as can be seen from a simple physical argument: Since the central spin is coupled to $N_b$ bath spins, for $S^z_{\rm tot}=\Or(1)$ it experiences an effective field $h_0/N_b$, and so does each bath spin. Thus the leading $h_0$-dependent part of the spin-spin correlation function scales as $\sim h_0^2/N_b^2$, which yields a contribution $\sim h_0^2/N_b$ to the energy. For $S^z_{\rm tot}=\Or(1)$ this is just the leading contribution from $\gamma_1^{(0)}+h_0$. For a small central field, this has been demonstrated in Eqs.~\refeq{gamma1}, \refeq{c1} and \refeq{cc2}.  

Since $\gamma_1$ is small compared to $\omega_0=\Or(N_b)$, we can still use Eq.~\refeq{exex} to determine $\o_0$ iteratively. Thus Eq.~\refeq{ensmallh0} is still applicable for the energy, resulting now in 
\be
E_0&=&-\frac{1}{4d} +\frac{h_0}{2}+\frac{\g_1^{(0)}}{2}- \frac{y_1^{(0)}}{2} -\frac{d}{2}\l[y_1^{(1)}-\l[y_1^{(0)}\r]^2+y_2^{(0)}+h_0 y_1^{(0)}\r]+ \Or(d^2)\,\label{e0classqm}\\
&=&-\sum_{j=1}^{N_b}\frac{\sqrt{1+\delta^2} A_j^2}{4\sqrt{(\nu-A_j)^2+(A_j \,\delta)^2}}- \frac{y_1^{(0)}}{2} -\frac{d}{2}\l[y_1^{(1)}-\l[y_1^{(0)}\r]^2+y_2^{(0)}+h_0 y_1^{(0)}\r]\label{e0explicit}
\ee
In the small-field limit, the results of the previous section are recovered. 

Comparing Eq.~\refeq{e0explicit} with Eq.~\refeq{hcl}, one identifies the leading classical contribution due to $h_0$ from Eq.~\refeq{hcl} stemming from the roots on the arc, i.e.~$\gamma_1^{(0)}$. The root $\omega_0$ encodes additional quantum-mechanical fluctuations which are of the same order of magnitude as the classical $h_0$-terms. 

In analogy to the energy, quantum fluctuations are also present in the correlation functions. In leading order, $\langle S_j^z\rangle$ is given by the classical expressions \refeq{m0} and \refeq{mj}. Fluctuations are due to $\omega_0$, which would yield a contribution $\sim d^2$ to $\langle S_j^z\rangle$, as in Eq.~\refeq{sjzh0pol}. However, the situation is different for the two-point function $\langle \vec S_0\cdot \vec S_j\rangle$: The quantum fluctuations in the energy lead to contributions of order $\Or(d)$ in the two-point function, cf.~Eq.~\refeq{sosjh0}. Taking together Eqs.~\refeq{sosj_class} and \refeq{sosjh0}, one obtains
\be
\langle \vec S_0\cdot \vec S_j\rangle_0=-\frac{(1+\delta^2)A_j-\nu}{4\sqrt{1+\delta^2}\sqrt{(\nu-A_j)^2+A_j^2\delta^2}}+\frac{d}{2} y_1^{(0)}-d A_j\label{sosj_class_qm}.
\ee
{Whereas Eq.~\refeq{sosjsmallh} is valid in the weak-field regime $h_0<d$ only, Eq.~\refeq{sosj_class_qm} gives the field dependence and the leading finite-size effects also for stronger fields $h_0>d$.} This result is compared to numerical data from complete diagonalization in Fig.~\ref{fig:sosj}, showing very good agreement. {Moreover, from Eq.~\refeq{sosj_class_qm}, it is clear how to separate classical from quantum fluctuations, giving nice insight into the essential physics of the model.}

\section{Conclusion}   
We have studied the exact solution of the central spin model, focussing on
spectral properties and static correlators. {In particular, it is possible to analyze the 
magnetization profile and the two-point correlation function using
a classical approximation, exact diagonalization, and the Bethe ansatz solution as three independent methods.}

{The exact magnetization profile of the quantum model 
follows the classical approximation very well already 
for small system sizes.  For a given distribution of coupling
parameters an increasing central field typically enhances the antiferromagnetic alignment 
of nearby spins, while it favours a ferromagnetic alignment with the outer spin.
The total magnetization of the system is typically small.}

{For the two-point correlation function a similar tendency can be observed, but 
the classical solution must be significantly corrected by quantum fluctuation terms
as given in Eq.~\refeq{sosj_class_qm}.   Only for the outermost spins the classical
solution tends to become exact.  The reason for this is that
in all cases we considered, classical contributions are encoded by the moments $x_{-\ell}$ of the couplings, whereas quantum fluctuations are expressed
in terms of the moments $x_\ell$, $\ell>0$. This means that the outer region of the
quantum dot, where the nuclear spins are coupled weakly to the electron spin,
are governed by classical physics, whereas the inner region experiences
stronger quantum fluctuations, due to the larger spin-exchange. }

{The classical approach is
analogous to the original BCS mean field solution of the superconducting
state.  Typically the classical approximation works better for the BCS 
model since quantum, i.e.~finite-size contributions are
sub-leading compared to the mean-field solution, whereas in the central spin model both can be of the same
order in the central spin model, depending on the quantity under
consideration. The reason for this is that the pairing amplitude $g= \Or(1/N)$
in the BCS model, whereas the analogous parameter $h_0\equiv g^{-1}= \Or(1)$ in the
central spin model. In view of tunable interactions in ultracold gases, this
could lead to the possibility of a new pairing phase for attractive electrons
with fixed particle number, when the attraction $g$ is of order one.  }

{After having demonstrated how to obtain the classical contributions
from the exact quantum-mechanical solution, we must emphasize that if 
$h_0\neq 0$, the expectation value $\langle \Lambda|S_j^{x,y}|\Lambda\rangle$ vanishes for all eigenstates $\Lambda$.
This is necessarily so, since $\Lambda$ must have a definite magnetization $S^z_{\rm tot}$ unless 
there is an accidental degeneracy in the system.
For the equivalent BCS model this means that the BCS order parameter
$\langle c^\dagger_{j\down}c^\dagger_{j\up}\rangle$ is identically zero 
for finite quantum systems.
{\it Technically, the well-known spontaneous symmetry breaking can therefore only be realized 
in the thermodynamic limit in the BCS model, despite the fact that a description in 
terms of the mean field solution (i.e.~classical vectors $\vec m_j$) gives 
quantitatively good results also for finite systems. }
This is in contrast to the prototypical example for symmetry breaking in 
ferromagnets, where the ground state and excited states generically already 
carry a non-zero expectation 
value of the order parameter for finite system sizes.}


Our results are of direct importance for the study of non-equilibrium dynamics: The understanding of the magnetization profile of eigenstates allows to estimate overlaps of eigenstates with those non-eigenstates which are realistic initial states in the time-evolution of the electron coupled to the nuclear spins. The computation of those overlaps is crucial in order to estimate the decoherence time. We leave this as a promising route for future research here.  

More generally, the study of classical and quantum contributions during the time evolution of non-equilibrium dynamics remains an important open question for future research. 
 
\subsubsection*{Acknowledgment}
We are grateful to J.-S. Caux, F.H.L.~Essler, M.A.~Jivulescu, A.~Kl\"umper, Z.~Kurucz, F.~G\"ohmann, I.~Schneider and A.~Struck for
useful discussions. M.B.~thanks the Rudolf-Peierls-Centre for Theoretical
Physics, University of Oxford, for kind hospitality, where part of this work
has been carried out. Financial support by the European network INSTANS and
the SFB-TR49 is gratefully acknowledged. 

\end{document}